\documentclass[twocolumn]{article}
\usepackage{graphicx}
\usepackage{subcaption}
\usepackage[margin=1.8cm, bottom=2.0cm]{geometry}
\usepackage[T1]{fontenc}
\usepackage[utf8]{inputenc}
\usepackage{mathtools}
\usepackage{hyperref}
\usepackage{cleveref}
\usepackage{siunitx}
\usepackage{chemformula}

\usepackage[backend=biber,style=ieee,sorting=none,citestyle=numeric-comp,giveninits=true,date=year,backref=true,maxbibnames=5]{biblatex}
\DeclareFieldFormat
  [inproceedings,incollection, misc]
  {title}{\mkbibquote{#1\isdot}}

\DeclareFieldFormat
  [inproceedings,incollection, misc]
  {booktitle}{\textit{#1}}

\newcommand{\affiliation}[1]{
  \begingroup
  \centering
  \small#1\par
  \endgroup
}
\newcommand{\auth}[2][]{#2\textsuperscript{#1}}
\newcommand{\Rsq}{\ensuremath{R^2}}
\newcommand{\simi}{\ensuremath{\mathord{\sim}}}

\begin{document}

\title{\textbf{Matbench Discovery}\\A framework to evaluate machine learning crystal stability predictions}

\author{
  \auth[1,2]{Janosh Riebesell}\qquad
  \auth[1]{Rhys E. A. Goodall}\qquad
  \auth[3]{Philipp Benner}\qquad
  \\[1ex]
  \auth[2,4]{Yuan Chiang}\qquad
  \auth[2,4]{Bowen Deng}\qquad
  \auth[2,4]{Gerbrand Ceder}\qquad
  \auth[2,4]{Mark Asta}\qquad
  \\[1ex]
  \auth[1]{Alpha A. Lee}\qquad
  \auth[2]{Anubhav Jain}\qquad
  \auth[2,4]{Kristin A. Persson}
}
\twocolumn[
  \begin{@twocolumnfalse}
    \maketitle
    \affiliation{1 Department of Physics, University of Cambridge, UK}
    \affiliation{2 Lawrence Berkeley National Laboratory, USA}
    \affiliation{3 Federal Institute of Materials Research and Testing (BAM), Germany}
    \affiliation{4 Department of Materials Science and Engineering, University of California - Berkeley, USA}

    \begin{abstract}
      \normalsize
      The rapid adoption of machine learning (ML) in domain sciences necessitates best practices and standardized benchmarking for performance evaluation. We present Matbench Discovery, an evaluation framework for ML energy models, applied as pre-filters for high-throughput searches of stable inorganic crystals. This framework addresses the disconnect between thermodynamic stability and formation energy, as well as retrospective vs. prospective benchmarking in materials discovery.
      We release a Python package to support model submissions and maintain an online leaderboard, offering insights into performance trade-offs. To identify the best-performing ML methodologies for materials discovery, we benchmarked various approaches, including random forests, graph neural networks (GNNs), one-shot predictors, iterative Bayesian optimizers, and universal interatomic potentials (UIP).
      Our initial results rank models by test set F1 scores for thermodynamic stability prediction: EquiformerV2 + DeNS > Orb > SevenNet > MACE > CHGNet > M3GNet > ALIGNN > MEGNet > CGCNN > CGCNN+P > Wrenformer > BOWSR > Voronoi fingerprint random forest. UIPs emerge as the top performers, achieving F1 scores of 0.57–0.82 and discovery acceleration factors (DAF) of up to 6x on the first 10k stable predictions compared to random selection.
      We also identify a misalignment between regression metrics and task-relevant classification metrics. Accurate regressors can yield high false-positive rates near the decision boundary at 0 eV/atom above the convex hull. Our results demonstrate UIPs’ ability to optimize computational budget allocation for expanding materials databases. However, their limitations remain underexplored in traditional benchmarks. We advocate for task-based evaluation frameworks, as implemented here, to address these limitations and advance ML-guided materials discovery.
    \end{abstract}
  \end{@twocolumnfalse}
  \vskip 0.5cm
]

\footnotetext{Correspondence to \href{mailto:janosh.riebesell@gmail.com}{janosh.riebesell@gmail.com, kristinpersson@berkeley.edu}}

\section{Introduction}
\label{sec:introduction}

The challenge of evaluating and benchmarking the rapid evolution of machine learning models is common to many domain sciences.  Specifically, the lack of pre-agreed upon tasks and data sets can obscure the performance of the model, making comparisons difficult. Material science presents one such domain, where in the last decade, the number of machine learning publications and associated models have increased dramatically. Similar to other domains, such as drug discovery and protein design, the ultimate success is often associated with the discovery of a new material with specific functionality. In the combinatorial sense, material science can be viewed as an optimization problem of mixing and arranging different atoms with a merit function that captures the complex range of properties that emerge.
To date, $\simi 10^5$ combinations have been tested experimentally
\cite{bergerhoff_inorganic_1983, belsky_new_2002}
, $\simi 10^7$ have been simulated\cite{jain_commentary_2013, saal_materials_2013, curtarolo_aflow_2012, draxl_nomad_2018, schmidt2024improving}, and upwards of $\simi 10^{10}$ possible quaternary materials are allowed by electronegativity and charge-balancing rules \cite{davies_computational_2016}.
The space of quinternaries and higher is even less explored, leaving vast numbers of potentially useful materials to be discovered.
The discovery of new materials is a key driver of technological progress and lies on the path to more efficient solar cells, lighter and longer-lived batteries, smaller and more efficient transistor gates just to name a few.
In light of our sustainability goals, these advances cannot come fast enough. Any speed-up new discovery methods might yield should be leveraged to their fullest extent.


Despite significant advances in empirical, theoretical, and computational materials science, discovering new materials still requires complex calculations, labor-intensive trial-and-error experimentation, and often happens fortuitously rather than through rational design.
Machine learning (ML) methods efficiently extract and distill trends from large datasets, and can handle high dimensionality, multiple objectives \cite{riebesell_pushing_2024}, uncertainty \cite{borg_quantifying_2022,goodall_rapid_2022,zhu_fast_2023}, and noisy or sparse data \cite{depeweg_decomposition_2017,bartel_critical_2020}, making them powerful additions to augment the traditional computational materials science tool set.

In particular, we focus on the role of ML to accelerate the use of Kohn-Sham density-functional theory (DFT) in the materials discovery pipeline.
In comparison to other simulation frameworks DFT offers a compelling compromise between fidelity and cost that has seen it adopted as a workhorse method by the computational material science community.
The great strengths of DFT as a methodology have led it to demand up to 45\% of core hours at the UK-based Archer2 Tier 1 supercomputer \cite{ukriGoldilocksConvergence} and over 70\% allocation time in materials science sector at NERSC \cite{griffin2022computational, NERSC:2018}. This heavy resource requirement drives demand for ways to reduce or alleviate its computational burden, such as efficiency improvements or substitution from ML approaches.

ML models are less accurate and reliable but orders of magnitude faster than ab-initio simulation.
This makes them most suitable for use in high-throughput (HT) searches to triage more expensive, higher-fidelity simulation methods such as DFT.
The use of neural networks for learning the DFT potential energy surface (PES) can be traced as far back as \cite{behler_generalized_2007}.
This work kicked off rapid advances and significant efforts to fit ever more sophisticated ML models to known samples of the PES.
Initially, most of these models were trained and deployed as interatomic potentials (also known as force fields) to study known materials of interest, a workflow that requires curating custom training data for each new system of interest \cite{bartok_machine_2018, deringer_general-purpose_2020}.
As larger and more diverse datasets have emerged from initiatives like the Materials Project (MP) \cite{jain_commentary_2013}, AFLOW \cite{curtarolo_aflow_2012} or the Open Quantum Materials Database (OQMD) \cite{saal_materials_2013}, researchers have begun to train so-called universal models that cover 90 or more of the most-application relevant elements in the periodic table.
This opens up the prospect of ML-guided materials discovery to increase the hit rate of stable crystals and speed up DFT- and expert-driven searches.


Progress in ML for materials is often measured according to performance on standard benchmark datasets.
As ML models have grown in complexity and applicability, benchmark datasets need to grow with them to accurately measure their usefulness.
However, due to the rapid pace of the field and the variety of possible approaches for framing the discovery problem, no large-scale benchmark yet exists for measuring the ability of ML to accelerate materials discovery.
As a result, it is unclear which methodologies or models are best suited for this task.
Recent works focusing on prospective computational materials discovery have proposed strategies based on one-shot coordinate free predictors \cite{goodall_rapid_2022}, iterative Bayesian optimizers \cite{zuo_accelerating_2021}, and universal interatomic potentials (UIP) \cite{chen_universal_2022, deng_chgnet_2023, batatia_mace_2023}.
These papers deploy their respective models on specific systems and custom datasets to highlight certain strengths but have yet to be compared in a systematic standardized manner that allows them to be ranked by their ability to accelerate materials discovery.
Our work aims to identify the state-of-the-art (SOTA) model by proposing a novel evaluation framework that closely simulates a real-world discovery campaign guided by ML models.
Our findings show that UIPs outperform all other methodologies we tested both in accuracy and robustness.

We hope that creation of benchmarks following this framework create a pathway through which interdisciplinary researchers with limited material science backgrounds can contribute usefully to model architecture and methodology development on a relevant task and thereby aid progress in Material Science.

\section{Evaluation Framework for Materials Discovery}
\label{sec:evaluation-framework-for-materials-discovery}

In this work we proposed a benchmark task that addresses four central challenges that we believe to be requirements when seeking to justify experimental validation of ML predictions:

\begin{enumerate}
  \item
        \textbf{Prospective Benchmarking}: Idealized and overly simplified benchmarks can fail to reflect the real-world challenges a model faces when used in an actual discovery campaign.
        This can result in a disconnect between benchmark metrics and real-world performance.
        Possible reasons for this include choosing the wrong target \cite{bartel_critical_2020} or picking an unrepresentative training and test splits \cite{wu_moleculenet_2018, kpanou_robustness_2021}.
        For small datasets of materials properties, `Leave-Out' data splitting strategies are often used to assess model performance \cite{meredig2018can, cubuk2019screening, zahrt2020cautionary}. However, in our target domain large quantities of diverse data ($\simi 10^5$) are available and hence retrospective splitting strategies predicated on clustering can end up testing artificial or unrepresentative use cases. This encourages using new sources of prospectively generated test data to understand application performance. Adopting this principle, the intended discovery workflow should be used to generate the test data leading to a substantial but realistic covariate shift between the training and test distributions that gives a much better indicator likely of performance on additional application of the same discovery workflow.

        \textbf{Relevant Targets:}
        In the case of materials discovery, high-throughput DFT calculated formation energies - although widely used as regression targets - do not indicate thermodynamic stability or synthesizability.
        In contrast, the thermodynamic competition between one material and its competing phases in the same chemical system determines its thermodynamic stability, which can be quantified by the material's energy to the convex hull phase diagram.
        The distance to the convex hull is the leading order term and strongly predictive of (meta-)stability at standard conditions \cite{sun2016thermodynamic}. Whilst other factors such as finite temperature effects, vibrational, and configurational entropy can modify a material's stability under real-world conditions they are prohibitive to calculate in high-throughput workflows, leaving the distance to the convex hull as the best signal-vs-data-availability compromise to use as a target.

        Moreover, ML models relying on relaxed crystal structures as input render any discovery pipeline circular since obtaining relaxed structures requires computationally expensive DFT simulations, thereby depending on the very process we intend to accelerate.
  \item
        \textbf{Informative Metrics}: Global error metrics like $\text{MAE}$, $\text{RMSE}$ and \Rsq{} can often give practitioners a false sense of security about the model's reliability. For example, accurate regressors are susceptible to unexpectedly high false-positive rates if those nominally accurate predictions lie close to the decision boundary and the failed predictions can incur a high opportunity cost by wasting lab time and resources. Consequently, models should be measured on their ability to lead to correct decision making patterns not regression accuracy. One way to do this is by defining a selection criteria and then evaluating regression models primarily in terms of their classification performance.
  \item
        \textbf{Scalability}: Future materials discovery efforts are likely to target broad chemical spaces and large data regimes.
        Small benchmarks can lack chemical diversity, and obfuscate poor scaling relations or weak out-of-distribution (OOD) performance.
        For instance, random forests achieve excellent performance on small datasets but are typically outperformed by neural networks on large datasets due to the benefits of representation learning \cite{goodall_predicting_2020}.
        Whilst we propose that large training sets are necessary to adequately differentiate the ability of models to learn in the larger data regime, given the enormous size of the configurational space of materials yet to be explored we also propose that it is important to construct a task where the test set is larger than the training set to mimic true deployment at scale. No other inorganic materials benchmarks test the prospects of large-scale deployment in this manner.
\end{enumerate}

We highlight two specific benchmarking efforts that have partially addressed the above challenges: Matbench \cite{dunn_benchmarking_2020} and the Open Catalyst Project (OCP) \cite{chanussot_open_2021}.
Other valuable efforts such as MatSciML \cite{lee2023matsciml} and JARVIS-Leaderboard \cite{choudhary2024jarvis} aggregate a wide variety of material science related benchmark tasks, including from Matbench and OCP, but do not introduce distinct benchmarking design patterns to those seen in Matbench or the OCP.

By providing a standardized collection of 13 datasets ranging in size from \simi300 to \simi132,000 samples from both DFT and experimental sources, Matbench addresses the scalability challenge, highlighting how model performance changes as a function of data regime.
Matbench helped focus the field of ML for materials, increase comparability across papers and provide a quantitative measure of progress in the field.
Importantly, all tasks were exclusively concerned with the properties of known materials.
We believe a task that simulates a materials discovery campaign by requiring materials stability prediction from unrelaxed structures to be a missing piece here.

OCP is a large-scale initiative aimed at discovering substrate-adsorbate combinations that can catalyze critical industrial reactions, transforming these adsorbates into more useful products.
The OCP has released two datasets thus far, OCP20 \cite{chanussot_open_2021} and OCP22 \cite{tran_open_2022} for training and benchmarking ML models.
OCP certainly addressed challenge 1 of closely mimicking a real-world problem.
They have recently shown that despite not reaching their target accuracy to entirely replace DFT, using ML in conjunction with confirmatory DFT calculations dramatically speeds up their combinatorial screening workflow \cite{lan_adsorbml_2023}.
The team behind the OCP has a second initiative targeting materials for direct air capture called OpenDAC that has shared the ODAC23 dataset \cite{sriram2024open}. The OpenDAC benchmark is setup identically to the OCP.

We believe that addressing these four challenges will result in benchmarks that enable future ML-guided discovery efforts to confidently select appropriate models and methodologies for the expansion of computational materials databases.

\begin{figure*}[t]
  \centering
  \includegraphics[width=0.75\linewidth]{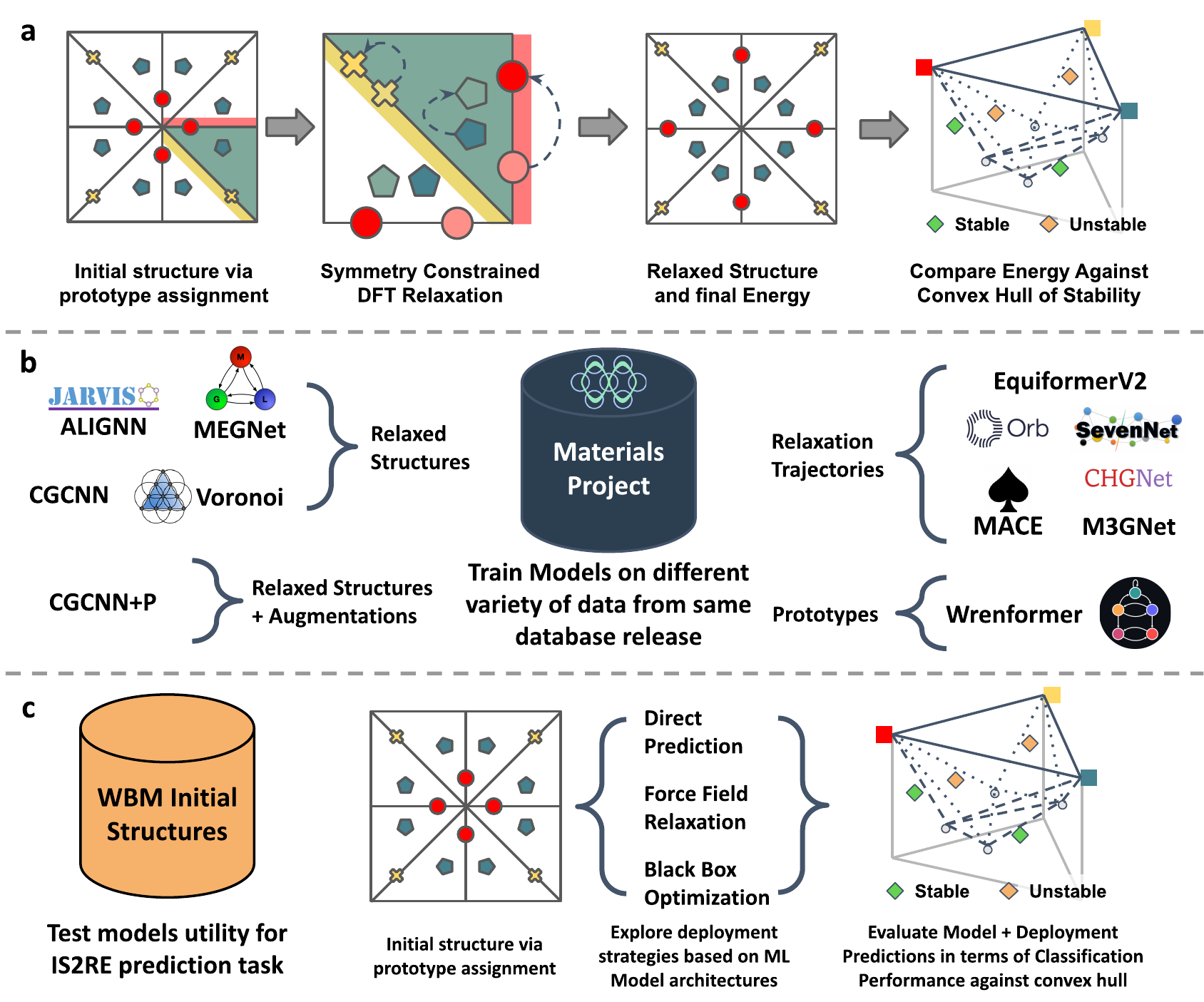}
  \caption{
    An overview of how data is used in Matbench-Discovery. a) shows a conventional prototype-based discovery workflow where different elemental assignments to the sites in a known prototype are used to create a candidate structure. This candidate is relaxed using DFT to arrive at a relaxed structure that can be compared against a reference convex hull. This sort of workflow was used to construct the WBM data set. b) highlights that databases such as the Materials Project provide a rich set of data which different academic groups have used to explore different types of models. While prior work tended to focus on individual modalities, our framework enables consistent model comparisons across modalities. c) shows the proposed test evaluation framework where the end user takes a machine learning model and uses it to predict a relaxed energy given an initial structure (IS2RE). This energy is then used to make a prediction as to whether the material will be stable or unstable with respect to a reference convex hull. From an applications perspective, this classification performance is better aligned with intended use cases in screening workflows.
    }
  \label{fig:overview}
\end{figure*}

\subsection{Matbench Discovery}

We propose a novel evaluation framework that places no constraints on the type of data a model is trained on as long as it would be available to a practitioner conducting a real materials discovery campaign.
This means that for the high-throughput DFT data considered, any subset of the energies, forces, stresses or any other properties that can be routinely extracted from DFT calculations, such as magnetic moments, are valid training targets.
All of these would be available to a practitioner performing a real materials discovery campaign and hence are permitted for training any model submission.
We only enforce that at test time, all models must make predictions on the convex hull distance of the relaxed structure with only the unrelaxed structure as input.
This setup avoids circularity in the discovery process, as unrelaxed structures can be cheaply enumerated through elemental substitution methodologies and do not contain information inaccessible in a prospective discovery campaign.
\autoref{fig:overview} provides a visual overview of design choices.

The convex hull distance of a relaxed structure is chosen as the measure of its thermodynamic stability, rather than the formation energy, as it informs the decision on whether to pursue a potential candidate crystal.
This decision was also motivated by \cite{bartel_critical_2020} who found that even composition-only models are capable of predicting DFT formation energies with useful accuracy.
However, when tasking those same models with predicting decomposition enthalpy, performance deteriorated sharply.
This insight exposes how ML models are much less useful than DFT for discovering new inorganic solids than would be expected given their low prediction errors for formation energies due to the impact of random as opposed to systematic errors.

Standard practice in ML benchmarks is to hold all variables fixed -- most importantly the training data -- and vary only the model architecture to isolate architectural effects on the performance.
We deliberately deviate from this practice due to diverging objectives from common ML benchmarks.
Our goal is to identify the best methodology for accelerating materials discovery.
What kind of training data a model can ingest is part of its methodology. Unlike energy-only models, UIPs benefit from the additional training data provided by the forces and stresses recorded in DFT relaxations. This allows them to learn a fundamentally higher fidelity model of the physical interactions between ions. That is a genuine advantage of the architecture and something any benchmark aiming to identify the optimal methodology for materials discovery must reflect.
In light of this utilitarian perspective, our benchmark contains models trained on varying datasets, and any model that can intake more physical modalities from DFT calculations is a valid model for materials discovery.

We define the Materials Project (MP) \cite{jain_commentary_2013} \href{https://docs.materialsproject.org/changes/database-versions}{v2022.10.28} database release as the maximum allowed training set for any compliant model submission.
Models may train on the complete set of relaxation frames, or any subset thereof such as the final relaxed structures.
Any subset of the energies, forces, and stresses are valid training targets.
In addition any auxiliary tasks such as predicting electron densities, magnetic moments, site-partitioned charges, etc. that can be extracted from the output of the DFT calculations are allowed for multi-task learning \cite{shoghi2023molecules}.
Our test set consists of the unrelaxed structures in the WBM dataset \cite{wang_predicting_2021}. Their target values are the PBE formation energies of the corresponding DFT-relaxed structures.

\subsubsection{Materials Project Training Set}
\label{sec:materials-project-training-set}

The Materials Project is a widely-used database of inorganic materials properties that have been calculated using high-throughput ab-initio methods.
At the time of writing, the Materials Project database \cite{jain_commentary_2013} has grown to \simi154 k crystals, covering diverse chemistries and providing relaxed and initial structures as well as the relaxation trajectory for every entry.

Our benchmark defines the training set as all data available from the \href{https://docs.materialsproject.org/changes/database-versions\#v2022.10.28}{v2022.10.28 MP release}.
We recorded a snapshot of energies, forces, stresses and magnetic moments for all MP ionic steps on 2023-03-15 as the canonical training set for Matbench Discovery, and provide convenience functions through our \href{https://pypi.org/project/matbench-discovery}{Python package} for easily feeding that data into future model submissions to our benchmark.

Flexibility in specifying the dataset allows authors to experiment with and fully exploit the available data.
This choice is motivated by two factors.
First, it may seem that models trained on multiple snapshots containing energies, forces, and stresses receive more training data than models trained only on the energies of relaxed structures. However, the critical factor is that all this additional data was generated as a byproduct of the workflow to produce relaxed structures. Consequently, all models are being trained using data acquired at the same overall cost.
If some architectures or approaches can leverage more of this byproduct data to make improved predictions this is a fair comparison between the two models.
This approach diverges philosophically from other benchmarks such as the OCP and Matbench where it has been more common to subcategorize different models and look at multiple alternative tasks (e.g. composition-only vs structure-available in Matbench or IS2RS, IS2RE, S2EF in OCP) and do not make direct comparisons of this manner.
Second, recent work in the space from \cite{li_critical_2023, li_exploiting_2023} has claimed that much of the data in large databases like MP are redundant and that models can be trained more effectively by taking a subset of these large data pools.
From a systems-level perspective, identifying novel cleaning or active-learning strategies to make better use of available data may be as crucial as architectural improvements, as both can similarly enhance performance, especially given the prevalence of errors in high-throughput DFT.
Consequently, such strategies where they lead to improved performance should be able to be recognized within the benchmark.
We encourage submissions to submit ablation studies showing how different system-level choices affect performance.
Another example of a system-level choice that may impact performance is the choice of optimizer, for example FIRE \cite{bitzek2006structural} vs L-BFGS, in the relaxation when using UIP models.

We highlight several example datasets that are valid within the rules of the benchmark that take advantage of these freedoms.
The first is the MP-crystals-2019.4.1 dataset \cite{chen_graph_2019} which is a subset of 133,420 crystals and their formation energies that form a subset of the v2021.02.08 MP release.
The MP-crystals-2022.10.28 dataset is introduced with this work comprising a set of 154,719 structures and their formation energies drawn from the v2021.02.08 MP release.
The next is the MPF.2021.2.8 dataset \cite{chen_universal_2022} curated to train the M3GNet model which takes a subset of 62,783 materials from the v2021.02.08 MP release.
The curators of the MPF.2021.2.8 dataset down-sampled the v2021.02.08 release significantly to select a subset of calculations that they believed to be most self-consistent.
Rather than taking every ionic step from the relaxation trajectory this dataset opts to select only the initial, final and one intermediate structure for each material to avoid biasing the dataset towards examples where more ionic steps were needed to relax the structure.
Consequently the dataset consists of 188,349 structures.
The MPF.2021.2.8 is a proper subset of the training data as no materials were deprecated between the v2021.02.08 and v2022.10.28 database releases.
The final dataset we highlight, with which several of the UIP models have been trained, is the MPtrj dataset \cite{deng_chgnet_2023}.
This dataset was curated from the earlier v2021.11.10 MP release. The MPtrj dataset is a proper subset of the allowed training data but several potentially anomalous examples from within MP were cleaned out of the dataset before the frames were subsampled to remove redundant frames.
It is worth noting that the v2022.10.28 release contains a small number of additional Perovskite structures not found in MPtrj that could be added to the training set within the scope of the benchmark.

We note that the v2023.11.1 deprecated a large number of calculations so data queried from subsequent database releases is not considered valid for this benchmark.

\subsubsection{WBM Test Set}
\label{sec:wbm-test-set}

The WBM dataset \cite{wang_predicting_2021} consists of 257,487 structures generated via chemical similarity-based elemental substitution of MP source structures followed by DFT relaxation and calculating each crystal's convex hull distance.
The element substitutions applied to a given source structure were determined by random sampling according to the weights in a chemical similarity matrix data-mined from the ICSD \cite{glawe_optimal_2016}.

The WBM authors performed 5 iterations of this substitution process (we refer to these steps as batches).
After each step, the newly generated structures found to be thermodynamically stable after DFT relaxation flow back into the source pool to partake in the next round of substitution.
This split of the data into batches of increasing substitution count is a unique and compelling feature of the test set as it allows out-of-distribution (OOD) testing by examining whether model performance degrades for later batches.
A higher number of elemental substitutions on average carries the structure further away from the region of material space covered by the MP training set (see \cref{fig:rolling-mae-vs-hull-dist-wbm-batches-models} for details).
Whilst this batch information makes the WBM dataset an exceptionally useful data source for examining the extrapolation performance of ML models, we look primarily at metrics that consider all batches as a single test set.

In order to control for the potential adverse effects of leakage between the MP training set and the WBM test set, we cleaned the WBM test set based on protostructure matching. We refer to the combination of a materials prototype and the elemental assignment of its wyckoff positions as a protostructure following \cite{parackal2024identifying}. First we removed 524 pathological structures in WBM based on formation energies being larger than 5 eV/atom or smaller than -5 eV/atom. We then removed from the WBM test set all examples where the final protostructure of a WBM material matched the final protostructure of an MP material. In total 11,175 materials were cleaned using this filter. We further removed all duplicated protostructures within WBM, keeping the lowest energy structure in each instance, leaving 215,488 structures in the unique prototype test set.

Throughout this work, we define stability as being on or below the convex hull of the MP training set ($E_\text{MP hull dist} \leq 0$).
32,942 out of 215,488 materials in WBM unique prototype test set satisfy this criterion.
Of these, \simi33k are unique prototypes, meaning they have no matching structure prototype in MP nor another higher-energy duplicate prototype in WBM.
Our code treats the stability threshold as a dynamic parameter, allowing for future model comparisons at different thresholds.
For initial analysis in this direction, see \cref{fig:roc-models} in the SI.

As WBM explores regions of materials space not well sampled by MP, many of the discovered materials that lie below MP's convex hull are not stable relative to each other.
Of the \simi33k that lie below the MP convex hull less than half, or around \simi20k, remain on the joint MP+WBM convex hull.
This observation suggests that many WBM structures are repeated samples in the same new chemical spaces.
It also highlights a critical aspect of this benchmark in that we knowingly operate on an incomplete convex hull.
Only current knowledge of competing points on the PES is accessible to a real discovery campaign and our metrics are designed to reflect this.

\subsubsection{Limitations of this Framework}

Whilst the framework proposed here directly mimics a common computational materials discovery workflow, it is worth highlighting that there still exist significant limitations to these traditional computational workflows that can prevent the material candidates suggested by such a workflow from being able to be synthesized in practice. For example, high-throughput DFT calculations often use small unit cells which can lead to artificial orderings of atoms. The corresponding real material may be disordered due to entropic effects that cannot be captured in the zero-kelvin thermodynamic convex hull approximated by DFT \cite{cheetham2024artificial}.

Another issue is that, when considering small unit cells, the DFT relaxations may get trapped at dynamically unstable saddle points in the true PES.
This failure can be detected by calculating the phonon spectra for materials predicted to be stable. However, the cost of doing so with DFT is often deemed prohibitive for high-throughput screening.
The lack of information about the dynamic stability of nominally stable materials in the WBM test set prevents this work from considering this important criteria as an additional screening filter.
However, recent progress in the development of UIPs suggests that ML approaches will soon provide sufficiently cheap approximations of these terms for high-throughput searches \cite{batatia2023foundation, deng2024overcoming}.
As the task presented here begins to saturate, we believe that future discovery benchmarks should extend upon the framework proposed here to also incorporate criteria based on dynamic stability.

When training UIP models there is a competition between how well given models can fit the energies, forces, and stresses simultaneously. The metrics in the Matbench Discovery leaderboard are skewed towards energies and consequently UIP models trained with higher weighting on energies can achieve better metrics. We caution that optimizing hyperparameters purely to improve performance on this benchmark may have unintended consequences for models intended for general purpose use. Practitioners should also consider other involved evaluation frameworks that explore orthogonal use cases when developing model architectures. We highlight work from \textcite{pota2024thermal} on thermal conductivity benchmarking, \textcite{fu2023forces} on MD stability for molecular simulation, and \textcite{chiang2024mlip} on modeling reactivity (hydrogen combustion) and asymptotic correctness (homonuclear diatomic energy curves) as complementary evaluation tasks for assessing the performance of UIP models.

We design the benchmark considering a positive label for classification as being on or below the convex hull of the MP training set. An alternative formulation would be to say that materials in WBM that are below the MP convex hull but do not sit on the combined MP+WBM convex hull are negatives. The issue with such a design is that it involves unstable evaluation metrics. If we consider the performance against the final combined convex hull rather than the initial MP convex hull, then each additional sample considered can retroactively change whether or not a previous candidate would be labeled as a success as it may no-longer sit on the hull. Since constructing the convex hull is computationally expensive, this path dependence makes it impractical to evaluate cumulative precision metrics (see \autoref{fig:cumulative-precision-recall}). The chosen setup does increase the number of positive labels and could consequently be interpreted as overestimating the performance. This overestimation decreases as the convex hull becomes better sampled. Future benchmarks building on this work could make use of combination of MP+WBM to control this artifact. An alternative framework could report metrics for each WBM batch in turn and retrain between batches, this approach was undesirable here as it increases the cost of submission five-fold and introduces many complexities, for example, should each model only retrain on candidates it believed to be positive, that would make fair comparison harder.

\subsection{Models}
\label{sec:models}

\begin{table*}[t]
  \centering
  \includegraphics[height=190pt]{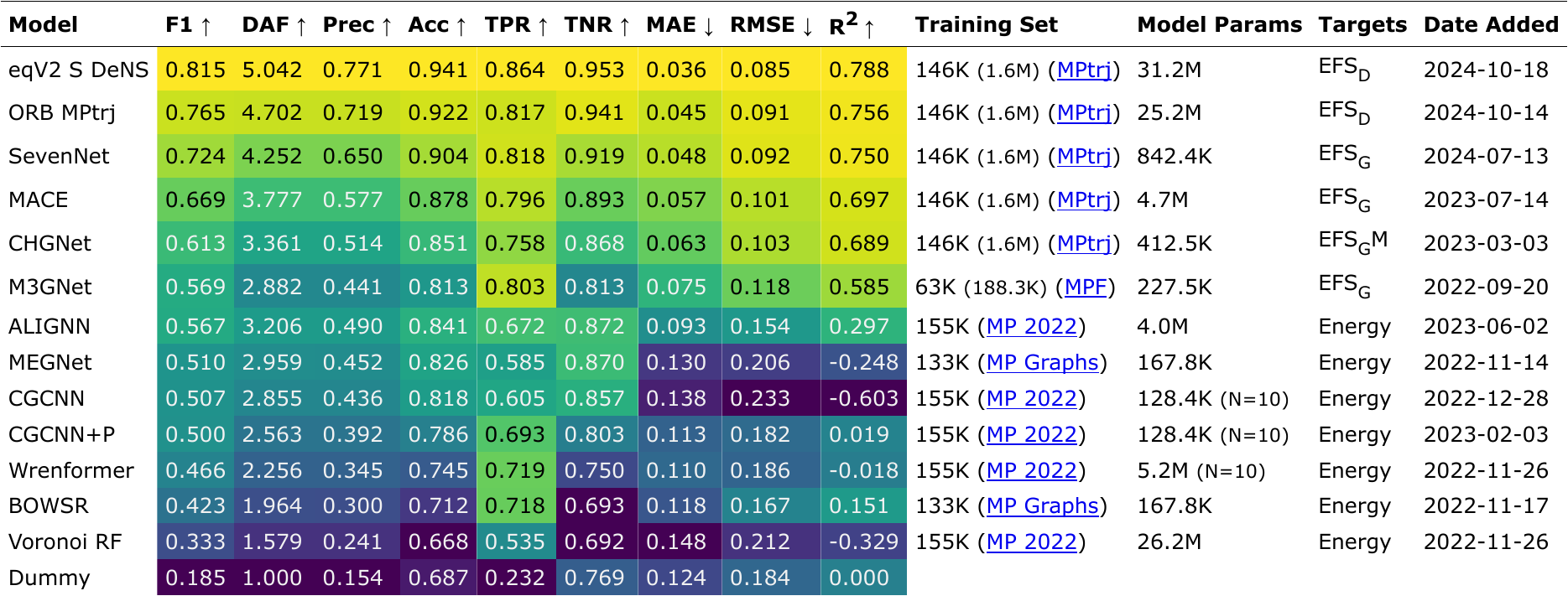}
  \caption{
    Classification and regression metrics for all models tested on our benchmark ranked by F1 score.
    The heat map ranges from yellow (best) to blue (worst) performance.
    DAF = discovery acceleration factor is the ratio of model precision to percentage of stable structures in the test set, TPR = true positive rate, TNR = true negative rate, MAE = mean absolute error, RMSE = root mean squared error.
    The dummy classifier uses the \texttt{scikit-learn} \texttt{stratified} strategy of randomly assigning stable or unstable labels according to the training set prevalence.
    The dummy regression metrics MAE, RMSE and \Rsq{} are attained by always predicting the test set mean.
    The top positions in the leaderboard are all taken by universal interatomic potential models trained on the combination of energies, forces and stresses. There is a pronounced gap in the regression metrics between the UIP models and the 7 energy-only models. It is worth noting that CGCNN+P, Wrenformer, and BOWSR achieve lower regression metrics through their mitigation strategies for initial and relaxed-structure mismatch but ultimately these strategies did not improve their usefulness as measured by the F1 score and DAF.
    Voronoi RF, CGCNN and MEGNet perform worse than dummy in regression metrics but better than dummy on some classification metrics, demonstrating that regression metrics alone can be misleading.
  }
  \label{tab:metrics-table-uniq-protos}
\end{table*}

To test a wide variety of methodologies proposed for learning the potential energy landscape, our initial benchmark release includes 13 models.
Next to each model's name we give the training targets that were used: E - Energy, F - Forces, S - Stresses and M - Magnetic moments. The subscripts G and D refer to whether gradient-based or direct prediction methods were used to obtain force and stress predictions.

\begin{enumerate}
  \item
        \textbf{EquiformerV2 + DeNS} \cite{liao2024equiformerV2, liao2024generalizing} (EFS$_D$) - EquiformerV2 builds on the first Equiformer model \cite{liao2023equiformer} by replacing the $SO(3)$ convolutions with eSCN convolutions \cite{passaro2023escn} as well as a range of additional tweaks to make better use of the ability to scale to higher $L_\text{max}$ using eSCN convolutions. EquiformerV2 uses direct force prediction rather than taking the forces as the derivative of the energy predictions for computational efficiency. Here we take the pre-trained "eqV2 S DeNS" \cite{barroso2024open} trained on the MPTrj dataset. This model in addition to supervised training on energies, forces, and stresses makes use of a auxilary training task based on de-noising non-equilibrium structures \cite{liao2024generalizing}. We refer to this model as "EquiformerV2 + DeNS" in the text and "eqV2 S DeNS" in plots.
  \item
        \textbf{Orb} \cite{neumann2024orb} (EFS$_D$) - Orb is a lightweight model architecture developed to scale well for the simulation of large systems such as metal organic frameworks. Rather than constructing an architecture that is equivariant by default, Orb instead makes use of data augmentation during training to achieve approximate equivariance. This simplifies the architecture allowing for faster inference. We report results for the "orb-mptrj-only-v2" model which was pre-trained using a diffusion-based task on MPTrj before supervised training on the energies, forces and stresses in MPTrj. For simplicity we refer to this model as "ORB MPTrj".
  \item
        \textbf{SevenNet} \cite{park2024scalable} (EFS$_G$) - SevenNet emerged from an effort to improve the performance of message passing neural networks \cite{gilmer2017neural} when used to conduct large scale simulations involving that benefit from parallelism via spatial decomposition. Here we use the pre-trained "SevenNet-0\_11July2024" trained on the MPTrj dataset. The SevenNet-0 model is an equivariant architecture based on a NequIP \cite{batzner_equivariant_2022} architecture that mostly adopts the GNoME \cite{merchant2023scaling} hyper-parameters. SevenNet-0 differs from NequIP and GNoME by replacing the tensor product in the self-connection layer with a linear layer applied directly to the node features, this reduces the number of parameters from  16.24 million in GNoME to 0.84 million for SevenNet-0. For simplicity we refer to this model as "SevenNet".
  \item
        \textbf{MACE} \cite{batatia_mace_2023} (EFS$_G$) - MACE builds upon the recent advances \cite{thomas_tensor_2018, batzner_equivariant_2022} in equivariant neural network architectures by proposing an approach to computing high-body-order features efficiently via Atomic Cluster Expansion \cite{drautz_atomic_2019}.
        Unlike the other UIP models considered MACE was primarily developed for molecular dynamics of single material systems and not the universal use case studied here.
        The authors trained MACE on the MPTrj dataset, these models have been shared under the name "MACE-MP-0" \cite{batatia2023foundation} and we report results for the "2023-12-03" version commonly called "MACE-MP-0 (medium)". For simplicity we refer to this model as "MACE".
  \item
        \textbf{CHGNet} \cite{deng_chgnet_2023} (EFS$_G$M) - CHGNet is a UIP for charge-informed atomistic modeling.
        Its distinguishing feature is that it was trained to predict magnetic moments on top of energies, forces and stresses in the MPtrj dataset (which was prepared for the purposes of training CHGNet).
        By modeling magnetic moments, CHGNet learns to accurately represent the orbital occupancy of electrons which allows it to predict both atomic and electronic degrees of freedom.
        We make use of the pre-trained "v0.3.0" CHGNet model from \cite{deng_chgnet_2023}.
  \item
        \textbf{M3GNet} \cite{chen_universal_2022} (EFS$_G$) - M3GNet is a GNN-based UIP for materials trained on up to 3-body interactions in the initial, middle and final frame of MP DFT relaxations.
        The model takes the unrelaxed input and emulates structure relaxation before predicting energy for the pseudo-relaxed structure.
        We make use of the pre-trained "v2022.9.20" M3GNet model from \cite{chen_universal_2022} trained on the compliant MPF.2021.2.8 dataset.
  \item
        \textbf{ALIGNN} \cite{choudhary_atomistic_2021} (E) - The Atomistic Line Graph Neural Network (ALIGNN) is a message passing GNN architecture that takes as input both the interatomic bond graph and a line graph corresponding to 3-body bond angles.
        The ALIGNN architecture involves a global pooling operation which means that it is ill-suited to force-field applications.
        To address this the ALIGNN-FF model was later introduced without global pooling \cite{choudhary_unified_2023}. We trained ALIGNN on the MP-crystals-2022.10.28 dataset for this benchmark.
  \item
        \textbf{MEGNet} \cite{chen_graph_2019} (E) - MatErials Graph Network is another GNN-based architecture that also updates a set of edge and global features (like pressure and temperature) in its message passing operation.
        This work showed that learned element embeddings encode periodic chemical trends and can be transfer-learned from large datasets (formation energies) to predictions on small data properties (band gaps, elastic moduli).
        We make use of the pre-trained "Eform\_MP\_2019" MEGNet model trained on the compliant MP-crystals-2019.4.1 dataset.
  \item
        \textbf{CGCNN} \cite{xie_crystal_2018} (E) - The Crystal Graph Convolutional Neural Network (CGCNN) was the first neural network model to directly learn 8 different DFT-computed material properties from a graph representing the atoms and bonds in a periodic crystal.
        CGCNN was among the first to show that just like in other areas of ML, given large enough training sets, neural networks can learn embeddings that outperform human-engineered structure features directly from the data.
        We trained an ensemble of 10 CGCNN models on the MP-crystals-2022.10.28 dataset for this benchmark.
  \item
        \textbf{CGCNN+P} \cite{gibson_data-augmentation_2022} (E) - This work proposes simple, physically motivated structure perturbations to augment stock CGCNN's training data of relaxed structures with structures resembling unrelaxed ones but mapped to the same DFT final energy.
        Here we chose $P=5$, meaning the training set is augmented with 5 random perturbations of each relaxed MP structure mapped to the same target energy.
        In contrast to all other structure-based GNNs considered in this benchmark, CGCNN+P is not attempting to learn the Born-Oppenheimer potential energy surface.
        The model is instead taught the PES as a step-function that maps each valley to its local minimum.
        The idea is that during testing on unrelaxed structures, the model will predict the energy of the nearest basin in the PES.
        The authors confirm this by demonstrating a lowering of the energy error on unrelaxed structures.
        We trained an ensemble of 10 CGCNN+P models on the MP-crystals-2022.10.28 dataset for this benchmark.
  \item
        \textbf{Wrenformer} (E) - For this benchmark, we introduce Wrenformer which is a variation on the coordinate-free Wren model \cite{goodall_rapid_2022} constructed using standard QKV-self-attention blocks \cite{vaswani_attention_2017} in place of message-passing layers.
        This architectural adaptation reduces the memory usage allowing the architecture to scale to structures with greater than 16 Wyckoff positions.
        Like its predecessor, Wrenformer is a fast coordinate-free model aimed at accelerating screening campaigns where even the unrelaxed structure is a priori unknown \cite{parackal2024identifying}.
        The key idea is that by training on the coordinate anonymized Wyckoff positions (symmetry-related positions in the crystal structure), the model learns to distinguish polymorphs while maintaining discrete and computationally enumerable inputs.
        The central methodological benefit of an enumerable input is that it allows users to predict the energy of all possible combinations of spacegroup and Wyckoff positions for a given composition and maximum unit cell size.
        The lowest-ranked protostructures can then be fed into downstream analysis or modeling.
        We trained an ensemble of 10 Wrenformer models on the MP-crystals-2022.10.28 dataset for this benchmark.
  \item
        \textbf{BOWSR} \cite{zuo_accelerating_2021} (E) - BOWSR combines a symmetry-constrained Bayesian optimizer (BO) with a surrogate energy model to perform an iterative exploration-exploitation-based search of the potential energy landscape.
        Here we use the pre-trained "Eform\_MP\_2019" MEGNet model \cite{chen_graph_2019} for the energy model as proposed in the original work. The high sample count needed to explore the PES with BO makes this by far the most expensive model tested.
  \item
        \textbf{Voronoi RF} \cite{ward_including_2017} (E) - A random forest trained to map a combination of composition-based Magpie features \cite{ward_general-purpose_2016} and structure-based relaxation-robust Voronoi tessellation features (effective coordination numbers, structural heterogeneity, local environment properties, \ldots) to DFT formation energies.
        This fingerprint-based model predates most deep learning for materials but significantly improved over earlier fingerprint-based methods such as the Coulomb matrix \cite{rupp_fast_2012} and partial radial distribution function features \cite{schutt_how_2014}.
        It serves as a baseline model to see how much value the learned featurization of deep learning models can extract from the increasingly large corpus of available training data.
        We trained Voronoi RF on the MP-crystals-2022.10.28 dataset for this benchmark.
\end{enumerate}

\section{Results}
\label{sec:results}


\Cref{tab:metrics-table-uniq-protos} shows performance metrics for all models included in the initial release of Matbench Discovery reported on the unique protostructure subset.
EquiformerV2 + DeNS takes the top spot and emerges as the current SOTA for ML-guided materials discovery.
It outperforms the other models on all 9 reported metrics.
When computing metrics in the presence of missing values or obviously pathological predictions (error of 5 eV/atom or greater) we assign the dummy regression values and a negative classification prediction to these points.
The discovery acceleration factor (DAF) measures how many more stable structures a model found compared to the dummy discovery rate achieved by randomly selecting test set crystals.
Formally the DAF is the ratio of the precision to the prevalence.
The maximum possible DAF is the inverse of the prevalence which on our dataset is $(\text{33k} / \text{215k})^{-1} \approx 6.5$.
Thus the current SOTA of 5.04 achieved by EquiformerV2 + DeNS leaves room for improvement.
However, evaluating each model on the subset of 10k materials each model ranks as being most stable (see \cref{tab:metrics-table-first-10k}), we see an impressive DAF of 6.33 for EquiformerV2 + DeNS which is approaching optimal performance for this task.

We find a large performance gap between models that make energy-only predictions directly from unrelaxed inputs (MEGNet, Wrenformer, CGCNN, CGCNN+P, ALIGNN, Voronoi RF) compared to UIPs that learn from force and stress labels and are used to emulate DFT relaxation before predicting a final energy.
While the F1 scores and DAFs of energy-only models are surprisingly performant, their regression metrics such as \Rsq and RMSE are significantly worse.
Of the energy-only models, only ALIGNN, BOWSR and CGCNN+P achieve positive coefficients of determination, \Rsq.
Negative \Rsq{} means model predictions explain the observed variation in the data less than simply predicting the test set mean.
In other words, these models are not predictive in a global sense (across the full dataset range).
However, even models with negative \Rsq{} can be locally good in the positive and negative tails of the test set distribution.
They suffer most in the mode of the distribution near the stability threshold of \SI{0}{eV/atom} above the hull.
This reveals an important shortcoming of \Rsq{} as a metric for classification tasks like stability prediction.

\begin{figure*}[htbp!]
  \centering
  \includegraphics[width=\linewidth]{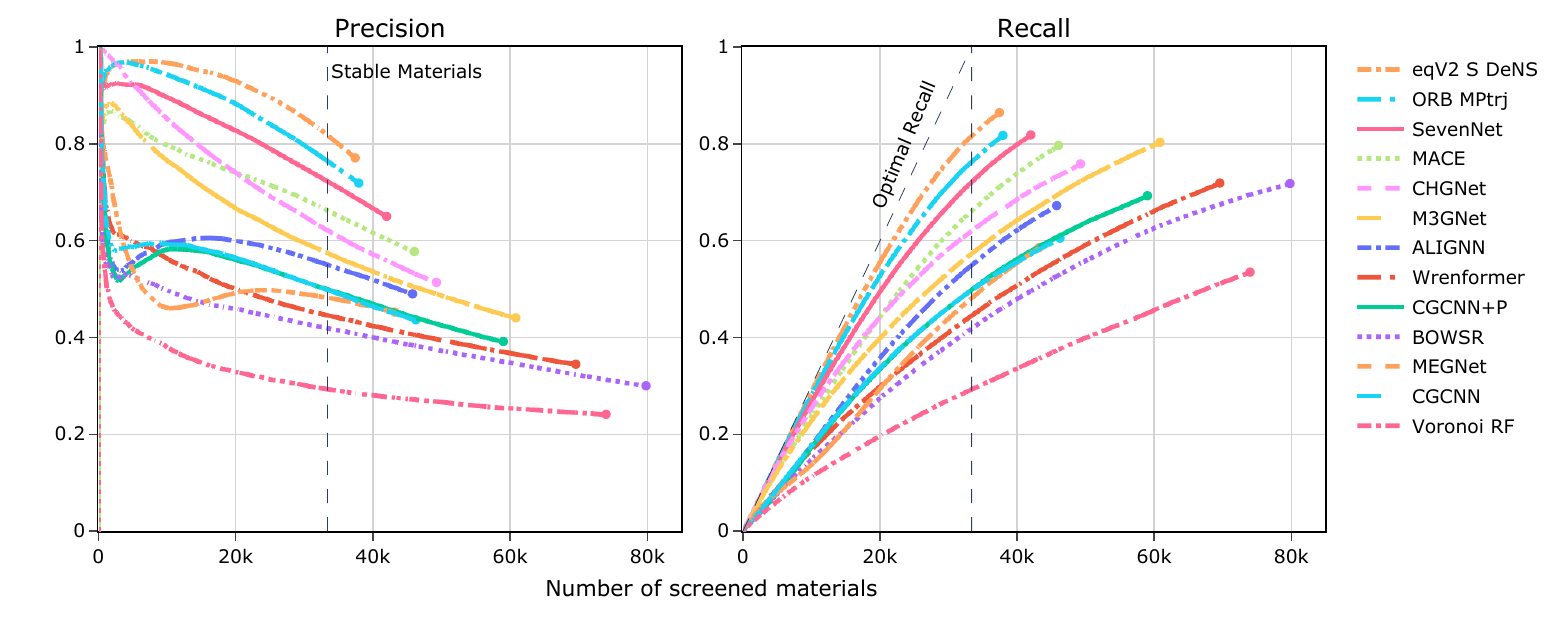}
  \caption{
    This figure measures model utility for materials discovery campaigns of varying sizes by plotting the precision and recall as a function of the number of model predictions validated.
    A typical discovery campaign will rank hypothetical materials by model-predicted hull distance from most to least stable and validate the most stable predictions first.
    A higher fraction of correct stable predictions corresponds to higher precision and fewer stable materials overlooked correspond to higher recall.
    Precision is calculated based only on the selected materials up to that point, whilst the cumulative recall depends on knowing the total number of positives upfront.
    This figure highlights how different models perform better or worse depending on the length of the discovery campaign.
    The UIPs are seen to offer significantly improved precision on shorter campaigns of \simi20k or less materials validated as they are less prone to false positive predictions among highly stable materials.
  }
  \label{fig:cumulative-precision-recall}
\end{figure*}

The reason CGCNN+P achieves better regression metrics than CGCNN but is still worse as a classifier becomes apparent from \cref{fig:hist-clf-pred-hull-dist-models} by noting that the CGCNN+P histogram is more sharply peaked at the 0 hull distance stability threshold.
This causes even small errors in the predicted convex hull distance to be large enough to invert a classification.
Again, this is evidence to choose carefully which metrics to optimize.
Regression metrics are far more prevalent when evaluating energy predictions. However, our benchmark treats energy predictions as merely means to an end to classify compound stability.
Improvements in regression accuracy are of limited use to materials discovery in their own right unless they also improve classification accuracy.
Our results demonstrate that this is not a given.


\Cref{fig:cumulative-precision-recall} has models rank materials by model-predicted hull distance from most to least stable; materials furthest below the known hull at the top, materials right on the hull at the bottom.
For each model, we iterate through that list and calculate at each step the precision and recall of correctly identified stable materials.
This simulates exactly how these models would be used in a prospective materials discovery campaign and reveals how a model's performance changes as a function of the discovery campaign length.
As a practitioner, you typically have a certain amount of resources available to validate model predictions.
These curves allow you to read off the best model given these constraints.
For instance, plotting the results in this manner shows that CHGNet initially achieves higher precision, than models such as EquiformerV2+DeNS, ORB MPTrj, SevenNet, and MACE that report higher precision across the whole test set.

In \Cref{fig:cumulative-precision-recall} each line terminates when the model believes there are no more materials in the WBM test set below the MP convex hull.
The dashed vertical line shows the actual number of stable structures in our test set.
All models are biased towards stability to some degree as they all overestimate this number, most of all BOWSR by 133\%.
This is only a problem in practice for exhaustive discovery campaigns that validate all stable predictions from a model.
More frequently, model predictions will be ranked most-to-least stable and validation stops after some pre-determined compute budget is spent, say, 10k DFT relaxations.
In that case, the concentration of false positive predictions that naturally accumulates near the less stable end of the candidate list can be avoided with no harm to the campaign's overall discovery rate (see \cref{tab:metrics-table-first-10k} where even the DAF of the worst performing model benchmarked, Voronoi RF, jumps from 1.58 to 2.49).

The diagonal `Optimal Recall' line on the recall plot in \Cref{fig:cumulative-precision-recall} would be achieved if a model never made a false negative prediction and stopped predicting stable crystals exactly when the true number of stable materials is reached.
Examining the UIP models, we find that they all achieve similar recall values, ranging from approximately 0.75 to 0.86.. This is substantially smaller than the variation we see in the precision for the same models \simi0.44-0.77. Inspecting the overlap, we find that the intersection of the models' correct agreements accounts for a precision of only 0.57 within the \simi0.75-0.86 range, with just 0.04 of the examples where all models are wrong simultaneously. These results indicate that the models are making meaningfully different predictions.

Examining the precision plot in \Cref{fig:cumulative-precision-recall}, we observe that the energy-only models exhibit a much more pronouced drop in their precision early-on, falling to 0.6 or less in the first 5k screened materials. Many of these models (all except BOWSR, Wrenformer and Voronoi RF) display an interesting hook shape in their cumulative precision, recovering again slightly in the middle of the simulated campaign between ~5k and up to ~30k before dropping again until the end.

\begin{figure}[t]
  \centering
  \includegraphics[width=\linewidth]{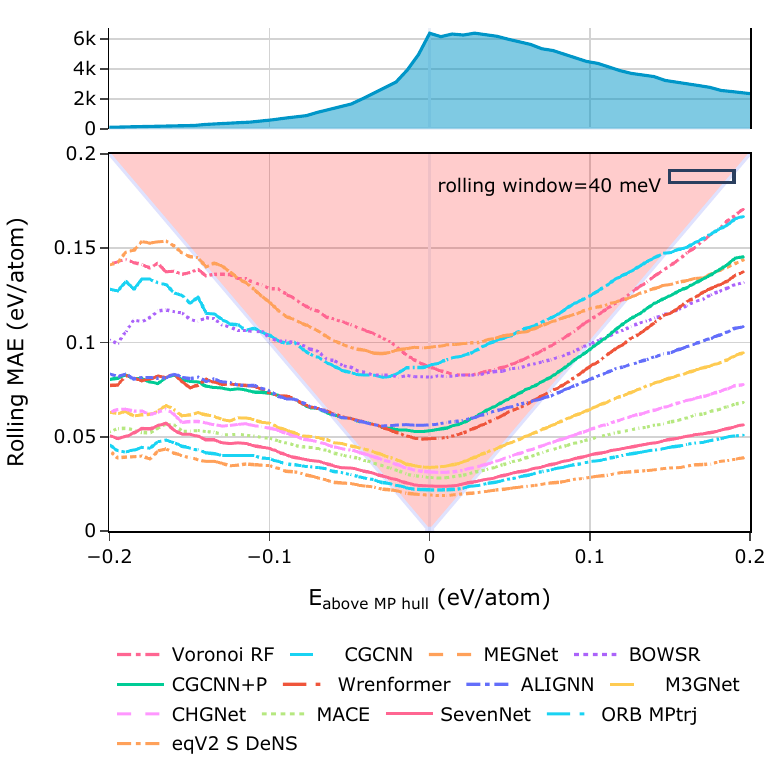}
  \caption{Universal potentials are more reliable classifiers because they exit the red triangle earliest.
    These lines show the rolling MAE on the WBM test set as the energy to the convex hull of the MP training set is varied.
    Lower is better.
    Inside the large red `triangle of peril', models are most likely to misclassify structures.
    As long as a model's rolling MAE remains inside the triangle, its mean error is larger than the distance to the convex hull.
    If the model's error for a given prediction happens to point towards the stability threshold at $E_\text{above MP hull} = 0$, its average error will change the stability classification from true positive or negative to false negative or positive.
    The width of the `rolling window' box indicates the width over which prediction errors were averaged.}
  \label{fig:rolling-mae-vs-hull-dist-models}
\end{figure}

\Cref{fig:rolling-mae-vs-hull-dist-models} provides a visual representation of the reliability of different models as a function of a material's DFT distance to the MP convex hull.
The lines show the rolling mean absolute error (RMAE) of model-predicted hull distances vs DFT.
The red-shaded area, which we coin the `triangle of peril', emphasizes the zone where the average model error surpasses the distance to the stability threshold at 0 eV.
As long as the rolling MAE remains within this triangle, the model is highly susceptible to misclassifying structures.
The average error in this region is larger than the distance to the classification threshold at 0, consequently in cases where the error points towards the stability threshold it would be large enough to flip a correct classification into an incorrect one. The sooner a model exits the triangle on the left side (negative DFT hull distance), the less likely it is to incorrectly predict stable structures as unstable, thereby reducing false negatives.
Exciting early on the right side (positive DFT hull distance) results in a lower likelihood of predicting unstable structures as stable, leading to fewer false positives.

All models tend to have lower rolling error towards the plot's left edge compared to the right edge.
This imbalance shows that models are more prone to false positive than false negative predictions. In other words, all models are less likely to predict a material at \SI{-0.2}{eV/atom} DFT hull distance as unstable than a material at \SI{0.2}{eV/atom} DFT hull distance as stable.
From a practitioner's standpoint, this is undesirable as attempting to validate an unstable material usually has a much higher opportunity cost than missing a stable one.
We hypothesize this error imbalance is due to the MP training set having an uncharacteristically high share of stable materials, causing statical models trained on it to be biased towards making low-energy predictions even for high-energy atomic configurations.
Training on datasets with more high-energy structures, such as Alexandria \cite{schmidt2024improving} and OMat24 \cite{barroso2024open}, would be expected to improve performance by balancing out this source of bias.

\section{Discussion}
\label{sec:discussion}

We have demonstrated the effectiveness of ML-based triage in high-throughput materials discovery and posit that the benefits of including ML in discovery workflows now clearly outweigh the costs.
\Cref{{tab:metrics-table-uniq-protos}} shows that in a realistic benchmark scenario that several models achieve a discovery acceleration greater than 2.5 across the whole dataset and up to 6 when considering only the 10k most stable predictions from each model (\cref{tab:metrics-table-first-10k}).
When starting this project, we were unsure which is the most promising ML methodology for high-throughput discovery.
Our findings demonstrate a clear superiority in the accuracy and extrapolation performance of UIPs.
Modeling forces enables these models to chart a path through atomic configuration space closer to the DFT-relaxed structure from where a more informed final energy prediction is possible.

As the convex hull becomes more comprehensively sampled through future discoveries, the fraction of unknown stable structures will naturally decline. This will lead to less enriched test sets and, consequently, more challenging and discriminative discovery benchmarks. However, the discovery task framed here addresses only a limited subset of potential UIP applications. We believe that additional benchmarks are essential to effectively guide UIP development. These efforts should prioritize task-based evaluation frameworks that address the four critical challenges we identify for narrowing the deployment gap: adopting prospective rather than retrospective benchmarking, tackling relevant targets, using informative metrics, and scalability.

Looking ahead, the consistently linear log-log learning curves observed in related literature \cite{vonlilienfeld_retrospective_2020} suggest that further decreases in the error of UIPs can be readily unlocked with increased training data. This has be borne out in the scaling results of GNoME \cite{merchant2023scaling}, MatterSim \cite{yang2024mattersim}, Alexandria \cite{schmidt2024improving}, and OMat24 \cite{barroso2024open} which all show improvements in performance when training on much larger datasets. To realize the full potential of scaling these models, future efforts should deploy their resources to generate large quantities of higher-than-PBE fidelity training data. The quality of a UIP model is circumscribed by the quality and level of theory of its training data.

Beyond simply predicting thermodynamic stability at zero Kelvin, future models will need to understand and predict material properties under varying environmental conditions, such as finite temperature and pressure, to aid in materials discovery. In this context, temperature-dependent dynamical properties constitute an area ripe for interatomic potentials. Another key open question is how effectively these models can contribute to the computational prediction of synthesis pathways.
Many existing approaches for reaction pathway prediction involve the use of heuristic rules for dealing with the significant added complexity of metastability alongside traditional ground state ab-initio data \cite{mcdermott_graph-based_2021, aykol_rational_2021, wen_chemical_2023}.
These algorithms will massively benefit from more efficient estimates of reaction energy barriers \cite{yuan2024analytical} and non-crystalline, out-of-equilibrium materials \cite{aykol_thermodynamic_2018}, opening up a whole new field to machine learning accelerated inquiry.

\section{Acknowledgments}
\label{sec:acknowledgments}

J.R. acknowledges support from the German Academic Scholarship Foundation (\href{https://wikipedia.org/wiki/Studienstiftung}{Studienstiftung}).
A.A.L. acknowledges support from the Royal Society.
A.J. and K.A.P. acknowledge the US Department of Energy, Office of Science, Office of Basic Energy Sciences, Materials Sciences and Engineering Division under contract no. DE-AC02-05-CH11231 (Materials Project program KC23MP).
This work used computational resources provided by the National Energy Research Scientific Computing Center (NERSC), a U.S. Department of Energy Office of Science User Facility operated under Contract No. DE-AC02-05CH11231.

Our profound gratitude extends to Hai-Chen Wang, Silvana Botti and Miguel A. L. Marques for their valuable contribution in crafting and freely sharing the WBM dataset.

We thank Rickard Armiento, Felix A. Faber and Abhijith S. Parackal for helping develop the evaluation procedures for Wren upon which this work builds. We also thank Rokas Elijosius for assisting in the initial implementation of Wrenformer and Mark Neumann, Luis Barroso-Luque and Yutack Park for submitting compliant models to the leaderboard.

We would like to thank Jason Blake Gibson, Shyue Ping Ong, Chi Chen, Tian Xie, Peichen Zhong and Ekin Dogus Cubuk for helpful discussions.

\section{Author Contributions}
\label{sec:author-contributions}

Janosh Riebesell: Methodology, Software, Data Curation, Investigation (Training: CGCNN, CGCCN+P, Wrenformer, Voronoi RF), Validation, Formal Analysis, Writing – original draft. Rhys Goodall: Conceptualization, Software, Validation, Formal Analysis, Writing – original draft, review \& editing. Philipp Benner: Software, Investigation (Training: ALIGNN, MACE), Writing – original draft. Yuan Chiang: Investigation (Training: MACE), Formal Analysis, Writing – review \& editing. Bowen Deng: Data Curation (MPtrj), Investigation (Training: CHGNet), Writing – review \& editing. Gerbrand Ceder: Supervision, Funding Acquisition. Mark Asta: Supervision, Funding Acquisition. Alpha Lee: Supervision. Anubhav Jain: Supervision. Kristin Persson: Supervision, Writing – review \& editing, Funding Acquisition.

\section{Code availability}
\label{sec:code-availability}

We welcome further model submissions to our GitHub repository \url{https://github.com/janosh/matbench-discovery}.

\section{Data availability}
\label{sec:data-availability}

We chose the latest Materials Project (MP) \cite{jain_commentary_2013} database release (\href{https://docs.materialsproject.org/changes/database-versions}{v2022.10.28} at time of writing) as the training set and the WBM dataset \cite{wang_predicting_2021} available at \url{https://figshare.com/articles/dataset/22715158} as the test set for this benchmark.
A snapshot of every ionic step including energies, forces, stresses and magnetic moments in the MP database is available at \url{https://figshare.com/articles/dataset/23713842}.

\clearpage
\sloppy
\printbibliography

\clearpage
\appendix

\section{Supplementary Information}
\label{sec:supplementary-information}

\setcounter{table}{0}
\setcounter{figure}{0}
\renewcommand{\thesection}{Note S\arabic{section}}
\renewcommand{\thefigure}{S\arabic{figure}}
\renewcommand{\thetable}{S\arabic{table}}

\subsection{Metrics on full test set and for 10k materials predicted most stable}
\label{sec:metrics-for-10k-materials-predicted-most-stable}

Unlike \cref{tab:metrics-table-uniq-protos} which evaluates model performance on the subset of unique WBM protostructures, \cref{tab:metrics-table} includes the full WBM test set of 257k materials.
The 44k additional materials in \cref{tab:metrics-table} excluded in \cref{tab:metrics-table-uniq-protos} comprise \num{11 175} discarded due to having a matching protostructure in MP plus another \num{32 784} materials which are protostructure duplicates of another WBM material with lower energy.
The most noteworthy difference between the two tables is a drop in DAF for all models in \cref{tab:metrics-table} compared to \cref{tab:metrics-table-uniq-protos}.
MACE for example achieves a DAF of 3.5 on the full test set (\cref{tab:metrics-table}) compared to 3.85 on the subset of 215.5k materials with unique protostructures (\cref{tab:metrics-table-uniq-protos}).
This \simi10\% increase is largely due to a \simi10\% decrease in the fraction of materials below the MP convex hull: 15.3\% (32,942 out of 215,488) in \cref{tab:metrics-table-uniq-protos} vs 16.7\% (42,825 out of 256,963) in the full dataset (\cref{tab:metrics-table}).
Since DAF is the ratio of the model's precision for stability prediction to the prevalence of stable structures, a lower prevalence results in a higher DAF.

While protostructures are non-trivial to match, thus potentially introducing bias into the test set by trying to deduplicate them, we still opted to feature \cref{tab:metrics-table-uniq-protos} in the main text since the removal of overlapping protostructures with MP makes it more closely reflect a model's true ability to extrapolate to out-of-domain materials.
Protostructure assignment is based on a combination of Aflow-style prototype labels and the chemical system \cite{hicks_aflow_2021} we use the \texttt{get\_protostructure\_label\_from\_spglib} implemented in \texttt{aviary} \cite{goodall2022aviary}. This string encodes the crystal's prototype and chemical system and is invariant to symmetry preserving changes to the structure. Consequently structures with different lattice parameters are flagged as protostructure duplicates as they would both be expected to relax to the same ground state in experimental conditions.

\begin{table*}
  \centering
  \includegraphics[height=190pt]{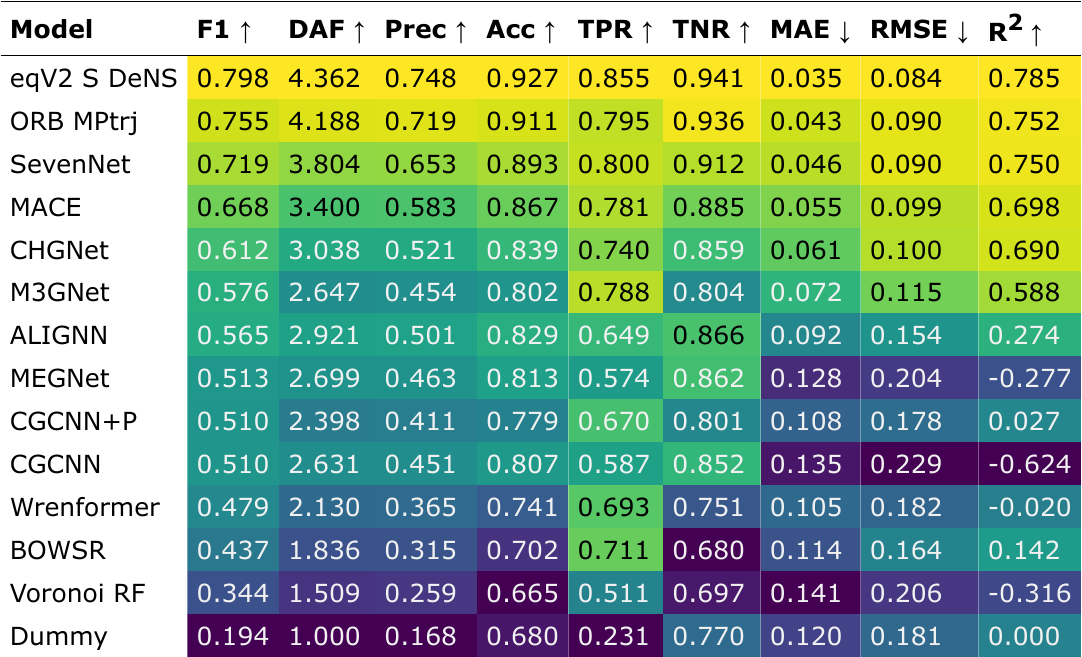}
  \caption{
    Same as \cref{tab:metrics-table-uniq-protos} but computed for all 257k structures in the WBM test set, no duplicate protostructures excluded.
    The prevalence of stable structures is 16.7\% (42,825 out of 256,963), placing a ceiling of 6 on the maximum possible DAF.
  }
  \label{tab:metrics-table}
\end{table*}

A real-world discovery campaign is unlikely to validate all stable predictions from a given model as we did in \cref{{tab:metrics-table-uniq-protos}}.
Presumably, it will rank model predictions from most to least stable and follow that list as far as time and compute budget permits.
Assuming that increases in compute resources will allow average future discovery campaigns to grow in scope, we believe 10k model validations to be a reasonable scope for average campaigns.
This is what \cref{tab:metrics-table-first-10k} simulates by calculating classification and regression metrics for the 10k test set materials predicted to be most stable by each model.
We again show dummy performance in the bottom row. Note that each model is now evaluated on a different slice of the data. However, the bottom row still shows dummy performance across the whole dataset.

\begin{table*}
  \centering
  \includegraphics[height=190pt]{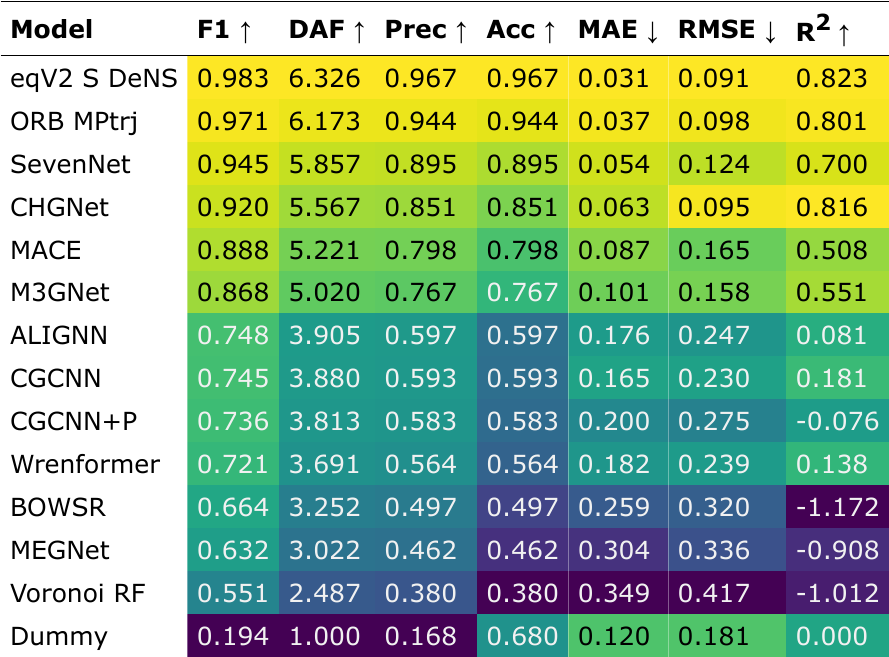}
  \caption{
    Stability prediction metrics for the 10k materials predicted to be most stable by each model.
    Every model picks out a different slice of the test set as the 10k materials it believes to be furthest below the known convex hull.
  }
  \label{tab:metrics-table-first-10k}
\end{table*}

\subsection{ROC Curves}
\label{sec:roc-curves}

\begin{figure*}
  \centering
  \includegraphics[width=0.7
\linewidth]{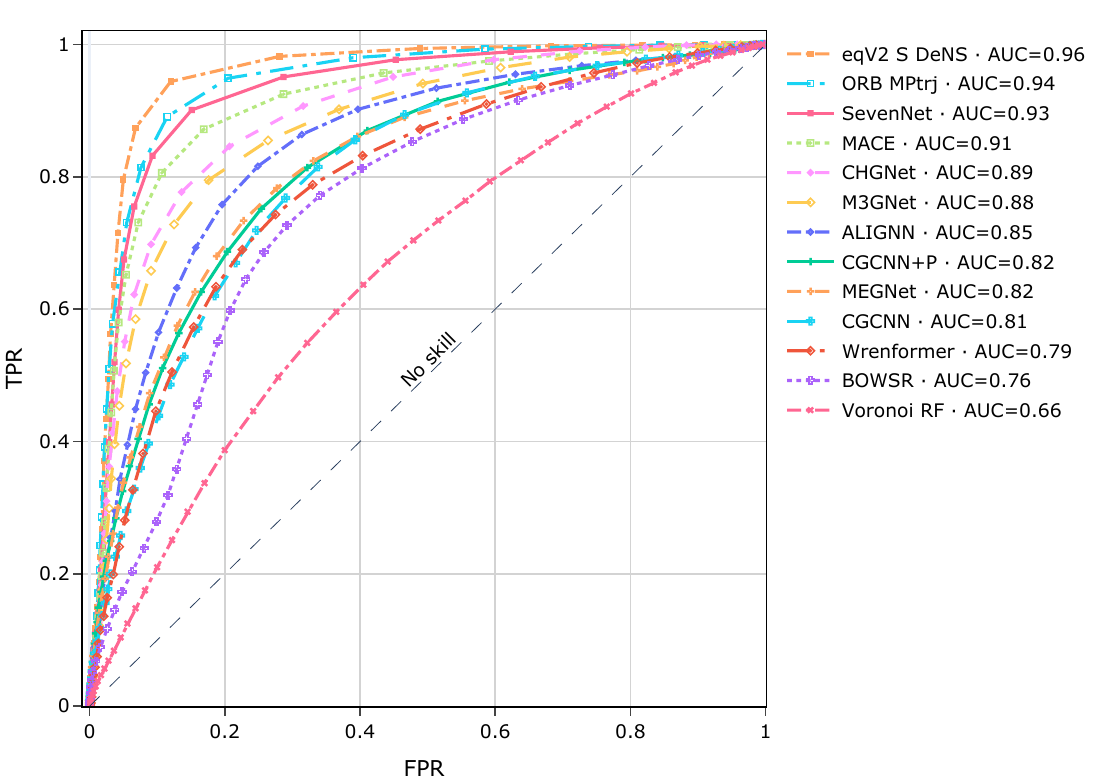}
  \caption{Receiver operating characteristic (ROC) curve for each model. The false positive rate (FPR) on the $x$ axis is the fraction of unstable structures classified as stable. The true positive rate (TPR) on the $y$ axis is the fraction of stable structures classified as stable.}
  \label{fig:roc-models}
\end{figure*}

A material is classified as stable if the predicted $E_\text{above hull}$ is below the stability threshold.
Since all models predict $E_\text{form}$, they are insensitive to changes in the threshold $t$.
The receiver operating characteristic (ROC) curve for each model is plotted in \cref{fig:roc-models}.
The diagonal `No skill' line shows the performance of a dummy model that randomly ranks material stability.

\subsection{Parity Plots}
\label{sec:parity-plots}

\begin{figure*}
  \centering
  \includegraphics[width=0.9\linewidth]{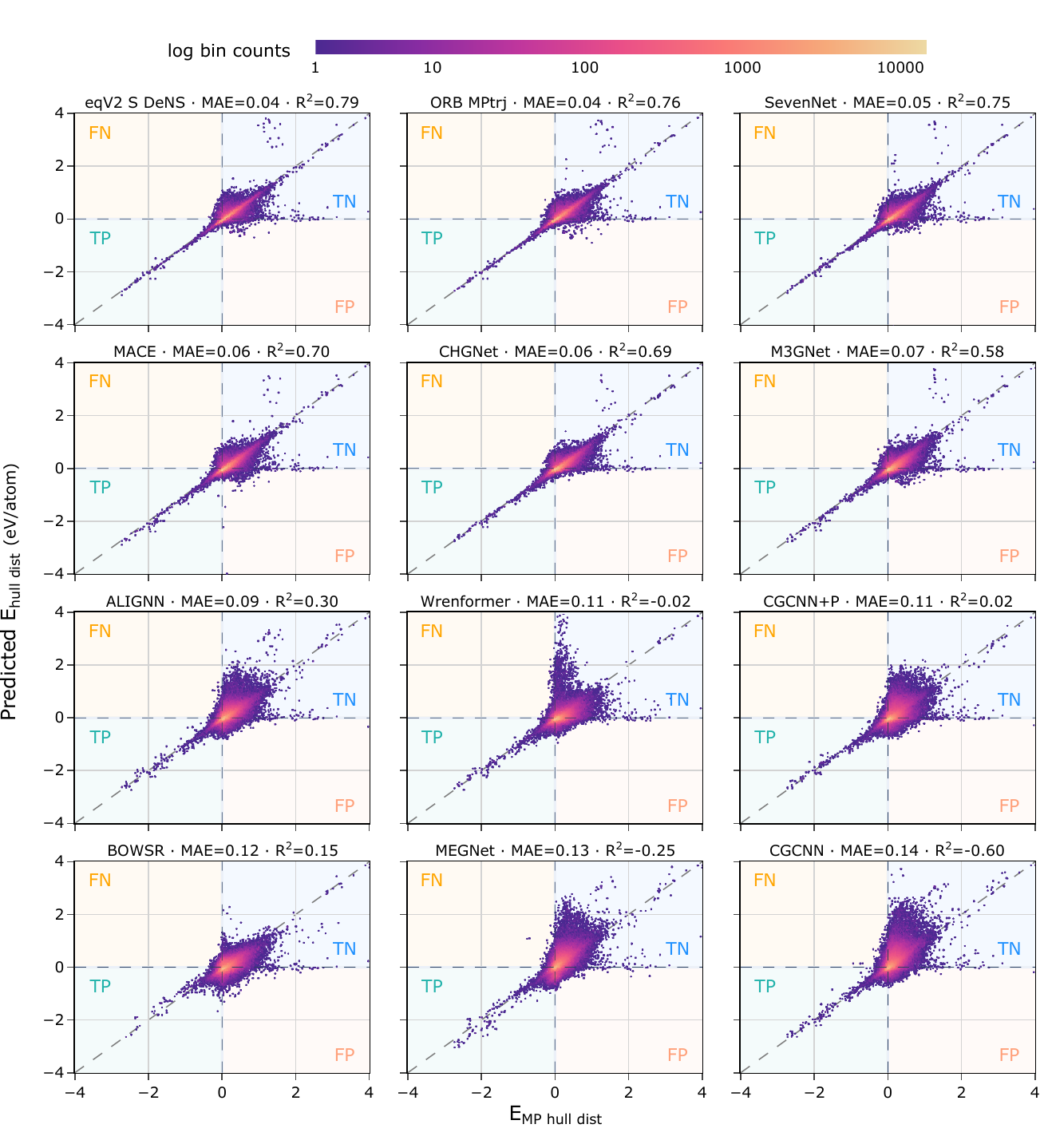}
  \caption{
    Parity plots of model-predicted energy distance to the convex hull (based on their formation energy predictions) vs DFT ground truth, color-coded by log density of points.
    Models are sorted left to right and top to bottom by MAE.
    For parity plots of formation energy predictions, see \cref{fig:e-form-parity-models}.
  }
  \label{fig:each-parity-models}
\end{figure*}

\begin{figure*}
  \centering
  \includegraphics[width=0.9\linewidth]{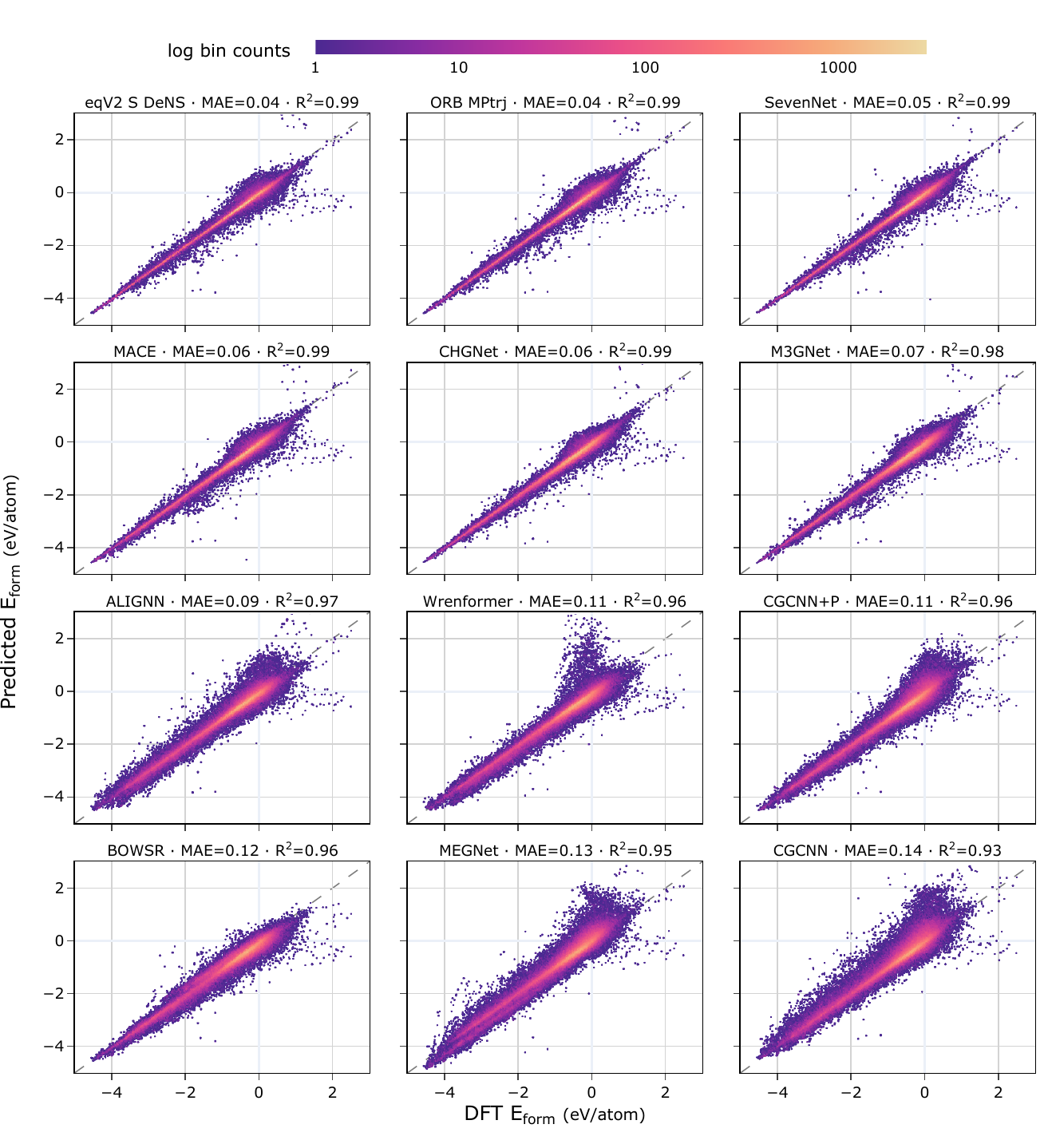}
  \caption{
    Parity plots of model-predicted formation energies vs DFT formation energies, color-coded by log density of points.
    The models are sorted from left to right and top to bottom by MAE.
    While similar to the parity plots in \cref{fig:each-parity-models} which shows the predicted distance to the convex hull vs DFT ground truth, this figure better visualizes the point density due to formation energy's wider spread.
    We observe broadly the same failure modes with occasional high DFT energy outliers predicted as near 0 formation energy by the models.
  }
  \label{fig:e-form-parity-models}
\end{figure*}

\Cref{fig:each-parity-models} shows that all models do well for materials far below the convex hull (left side of the plot). Performance for materials far above the convex hull is more varied with occasional underpredictions of the energy of materials far above the convex hull (right side). All models suffer most in the mode of the distribution at $x = 0$.

Two models stand out as anomalous to the general trends.

Wrenformer is the only model with a large number of severe energy overpredictions at $x = 0$ along the positive $y$ axis. We investigated these failure cases in more detail and found these overpredictions to be dominated by spacegroup 71 with poor representation in the training data. Digging into the MP-crystals-2022.10.28 dataset we see that on post analysis there are a large number of A$_2$BC compounds in spacegroup 71 in the v2022.10.28 that are trapped in local minima and thus much higher in energy. Given that Wrenformer nominally predicts the energy the global minima conditioned on the protostructure these are unsuitable for training Wrenformer and their inclusion leads to the errors seen on similar, correctly relaxed structures, in the WBM test set.

The other anomalous model is MACE with several severe underpredictions at $x = 0$ along the negative $y$ axis. We investigated these points for common traits in composition or crystal symmetry but noticed no pattern.

Beyond these MACE outliers visible in the plot, MACE exhibits another rare but reproducible type of failure case, in which the final predicted energy after relaxation is off by several orders of magnitude. The largest `derailed' prediction was $-10^{22}$ eV/atom for \texttt{wbm-3-31970} (formula H$_2$Ir). In each case, the MACE relaxation exhausted the maximum number of ionic steps set to 500 and caused a volume implosion from initial cell volumes of hundreds to relaxed cell volumes of tens of cubic Angstrom. Using the checkpoint trained on the M3GNet dataset which we received from the MACE authors, this failure mode occurred for several hundred of the 250k test set crystals. Using the checkpoint we trained ourselves on the MPtrj dataset, it affects only 44 test crystals, suggesting that these holes in the MACE PES can perhaps be fully plugged by further increasing the training set or even changing the loss function.
Further analysis is ongoing.
Since these derailed values are easily identified in practice when actually performing a prospective discovery campaign, we excluded them from the MACE parity plat and all other downstream analyses.

\subsection{Hull Distance Box plot}
\label{sec:hull-distance-box-plot}

\begin{figure*}
  \centering
  \includegraphics[width=0.9\linewidth]{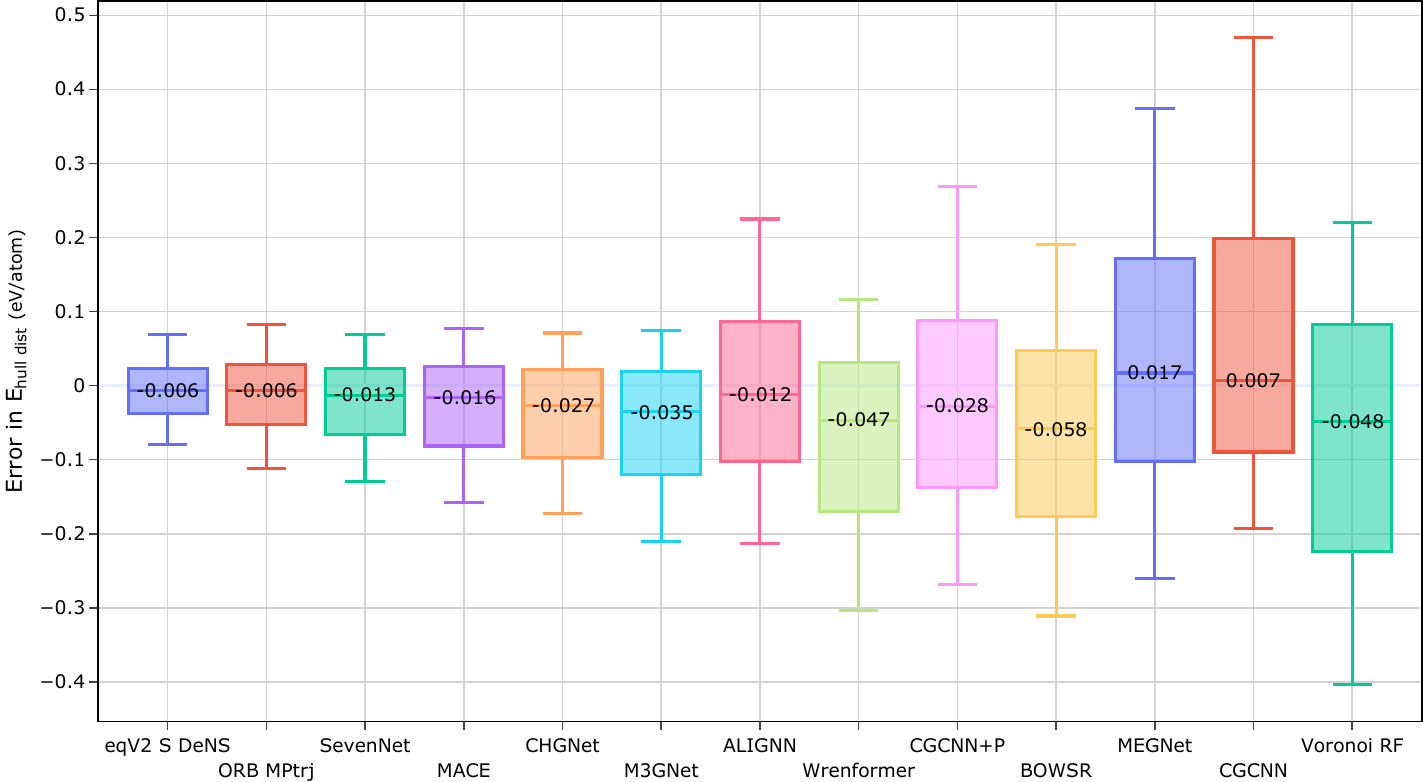}
  \caption{Box plot of interquartile ranges (IQR) of hull distance errors for each model. The whiskers extend to the 5th and 95th percentiles. The horizontal line inside the box shows the median. BOWSR has the highest median error, while Voronoi RF has the highest IQR.
  }
  \label{fig:box-hull-dist-errors}
\end{figure*}

\Cref{fig:box-hull-dist-errors} shows a box-plot of the errors in predicting the distance to the convex hull. The box demarks the interquartile range and the whiskers show the 5th and 95th percentiles.
BOWSR has the largest median error, while Voronoi RF has the largest IQR.
Looking at the models we see that there is a systematic trend to underpredict the distance to the convex hull.
This trend can be rationalized give the bias in the Materials Project towards stable and therefore low formation energy structures.
In contrast, due to it's construction the WBM test set is distributionally distinct with a higher average formation energy.
Therefore we propose that models trained on the Materials Project are likely result in such systematic underpredictions and this affect should be able to be addressed by training on additional data from a similar distribution to WBM i.e. Alexandria \cite{schmidt2024improving}.

\subsection{Classification Histograms using Model-Predicted Energies}
\label{sec:classification-histograms-using-model-predicted-energies}

\begin{figure*}
  \centering
  \includegraphics[width=0.9\linewidth]{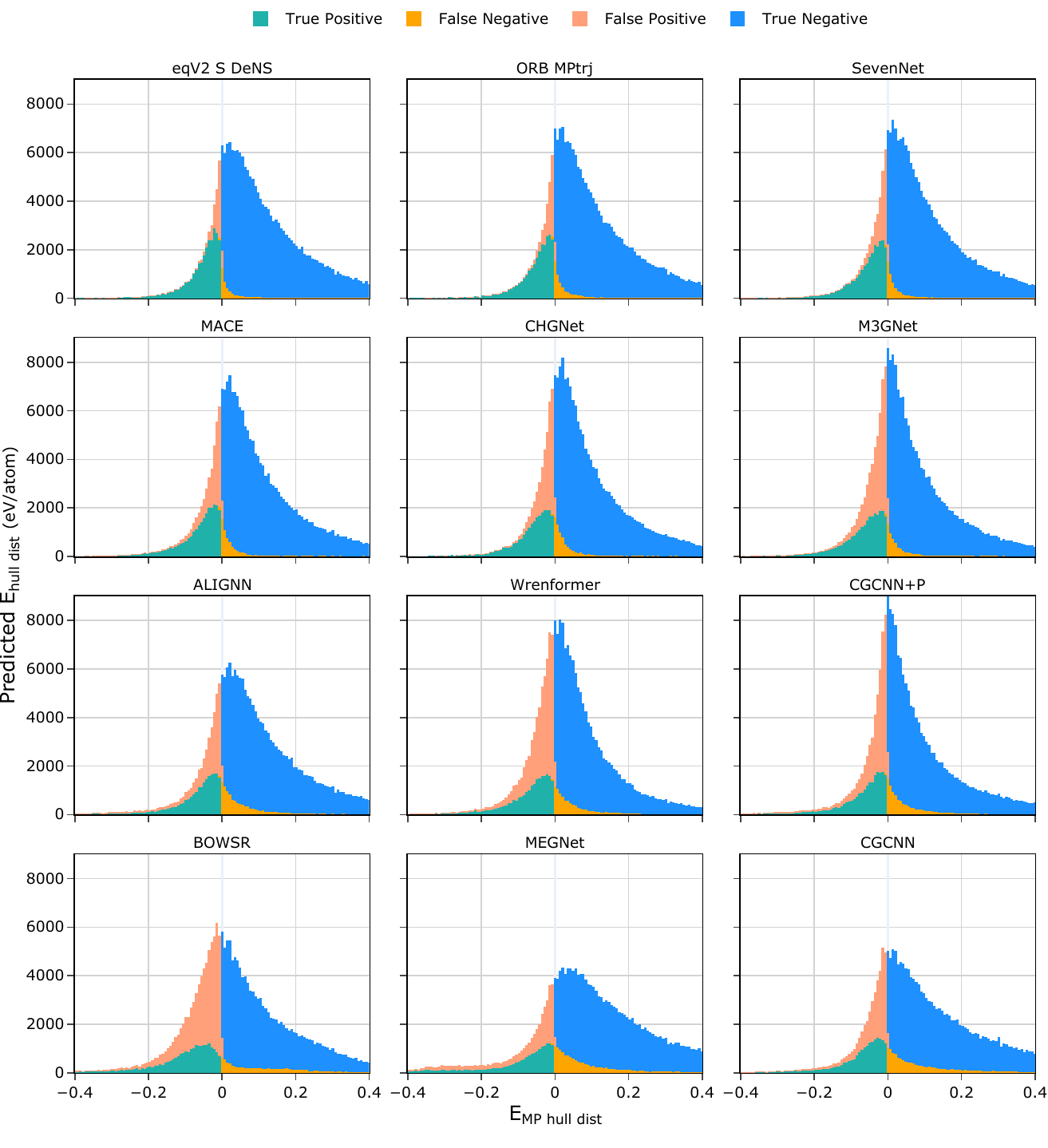}
  \caption{Distribution of model-predicted hull distance colored by stability classification. Models are sorted from top to bottom by F1 score. The thickness of the red and yellow bands shows how often models misclassify as a function of how far away from the convex hull they place a material.
  }
  \label{fig:hist-clf-pred-hull-dist-models}
\end{figure*}

\Cref{fig:hist-clf-pred-hull-dist-models} shows histograms of the model predicted distances to the convex hull colored by the stability classification.
These plots allow for practitioners to assess how the accuracy varies for materials depending on how far above or below the hull they are predicted to lie.
The most pronounced affect we see from the histograms is that the CGCNN+P histogram is more strongly peaked than CGCNN's giving much better agreement with the actual DFT ground truth distribution of hull distances for the test set. This explains why CGCNN+P performs better as a regressor, but also reveals how it can perform simultaneously worse as a classifier. By moving predictions closer to the stability threshold at \SI{0}{eV/atom} above the hull, even small errors are significant enough to tip a classification over the threshold.

\subsection{Measuring extrapolation performance from WBM batch robustness}

As a reminder, the WBM test set was generated in 5 successive batches, each step applying another element replacement to an MP source structure or a new stable crystal generated in one of the previous replacement rounds. The likelihood by which one element replaces another is governed by ISCD-mined chemical similarity scores for each pair of elements. Naively, one would expect model performance to degrade with increasing batch count, as repeated substitutions should on average `diffuse' deeper into uncharted regions of material space, requiring the model to extrapolate more. We observe this effect for some models much more than others.

\begin{figure*}[htbp!]
  \centering
  \includegraphics[width=0.9\linewidth]{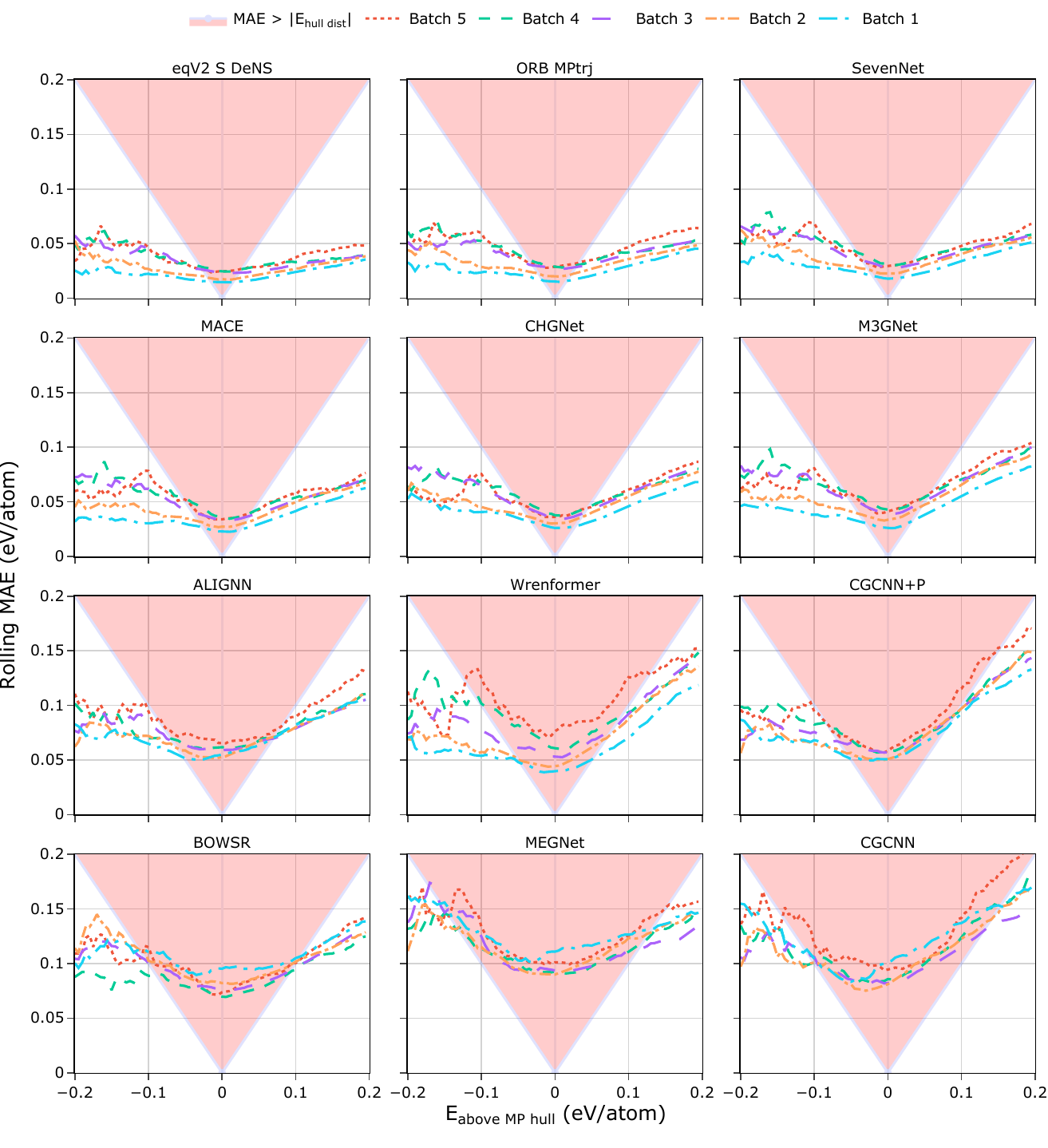}
  \caption{
    Rolling MAE as a function of distance to the convex hull for different models.
    Most models considered show a predictable decrease in performance from batch 1 to batch 5.
    The effect is larger for some models than others but batches 4 and 5 consistently incur the highest convex hull distance errors for all models.
    The UIPs show stronger extrapolative performance, as they show minimal deterioration in performance on later batches that move further away from the original MP training distribution.
    Wrenformer, by contrast, exhibits a pronounced increase in MAE with batch count.
    We view these plots as a strong indicator that Matbench Discovery is indeed testing out-of-distribution extrapolation performance as it is Occam's razor explanation for the observed model performance drop with increasing batch count.
  }
  \label{fig:rolling-mae-vs-hull-dist-wbm-batches-models}
\end{figure*}

\Cref{fig:rolling-mae-vs-hull-dist-wbm-batches-models} shows the rolling MAE as a function of distance to the convex hull for each of the 5 WBM rounds of elemental substitution. These plots show a stronger performance decrease for Wrenformer and Voronoi RF than for UIPs and even force-less GNNs with larger errors like ALIGNN, MEGNet and CGCNN.

\Cref{fig:rolling-mae-vs-hull-dist-wbm-batches-models} shows the rolling MAE for different models split out by calculation batch in the WBM data set.
MEGNet and CGCNN both incur a higher rolling MAE than Wrenformer across all 5 batches and across most or all of the hull distance range visible in these plots. However, similar to the UIPs, MEGNet and CGCNN exhibit very little degradation for higher batch counts. The fact that higher errors for later batches are specific to Wrenformer and Voronoi RF suggests that training only on composition and coarse-grained structural features (spacegroup and Wyckoff positions in the case of Wrenformer; coordination numbers, local environment properties, etc. in the case of Voronoi RF) alone is insufficient to learn an extrapolatable map of the PES.

Given its strong performance on batch 1, it is possible that given sufficiently diverse training data, Wrenformer could become similarly accurate to the UIPs across the whole PES landscape at substantially less training and inference cost. However, the loss of predictive forces and stresses may make Wrenformer unattractive for certain applications even then.

\subsection{Largest Errors vs.~DFT Hull Distance}
\label{sec:largest-errors-vs.-dft-hull-distance}

\begin{figure*}
  \centering
  \includegraphics[height=200pt]{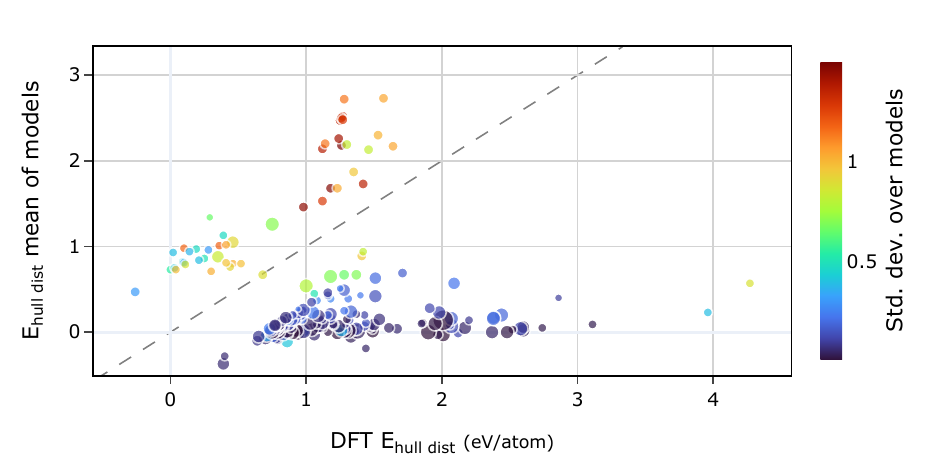}
  \caption{
    DFT vs predicted hull distance (average over all models) for the 200 largest error structures.
    Points are colored by model disagreement as measured by the standard deviation in hull distance predictions from different models.
    Point scale with the number of atoms in the structures.
    This plot shows that high-error predictions are biased towards predicting too small hull distances.
    This is unsurprising considering MP training data mainly consists of low-energy structures.
    There is a strong color separation between the mostly dark blue low-energy bias predictions and the high-error green or red predictions.
    Blue means models are in good agreement, i.e. all models are wrong together.
    Red means large-error predictions with little model agreement, i.e. all models are wrong in different ways.
    Some of the blue points with large errors yet good agreement among models may be accurate ML predictions for a DFT relaxation gone wrong.
    The dark blue points also tend to be larger corresponding to larger structures where DFT failures are less surprising.
    This suggests ML model committees might be used to cheaply screen large databases for DFT errors in a high-throughput manner.
  }
  \label{fig:scatter-largest-errors-models-mean-vs-true-hull-dist-all}
\end{figure*}

Given the large variety of models tested, we asked whether any additional insight into the errors can be gained from looking at how the predictions vary between different models.
In \cref{fig:scatter-largest-errors-models-mean-vs-true-hull-dist-all} we see two distinct groupings emerge when looking at the 200 structures with the largest errors.
This clustering is particularly apparent when points are colored by model disagreement.

\subsection{Exploratory Data Analysis}
\label{sec:eda}

\begin{figure*}
  \centering
  \begin{subfigure}[b]{0.9\linewidth}
    \includegraphics[width=\linewidth]{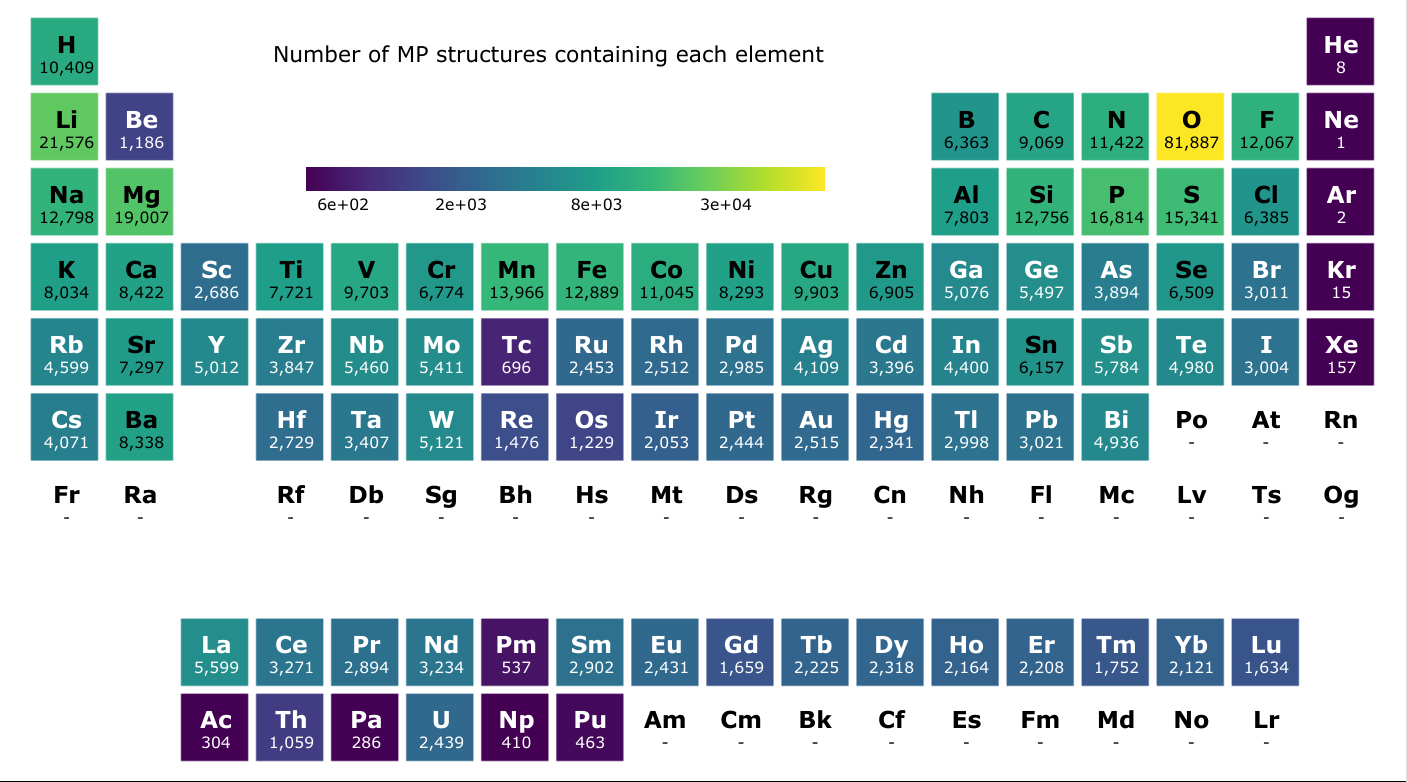}
    \caption{MP training set element occurrence}
    \label{fig:mp-element-counts-by-occurrence}
  \end{subfigure}
  \begin{subfigure}[b]{0.9\linewidth}
    \includegraphics[width=\linewidth]{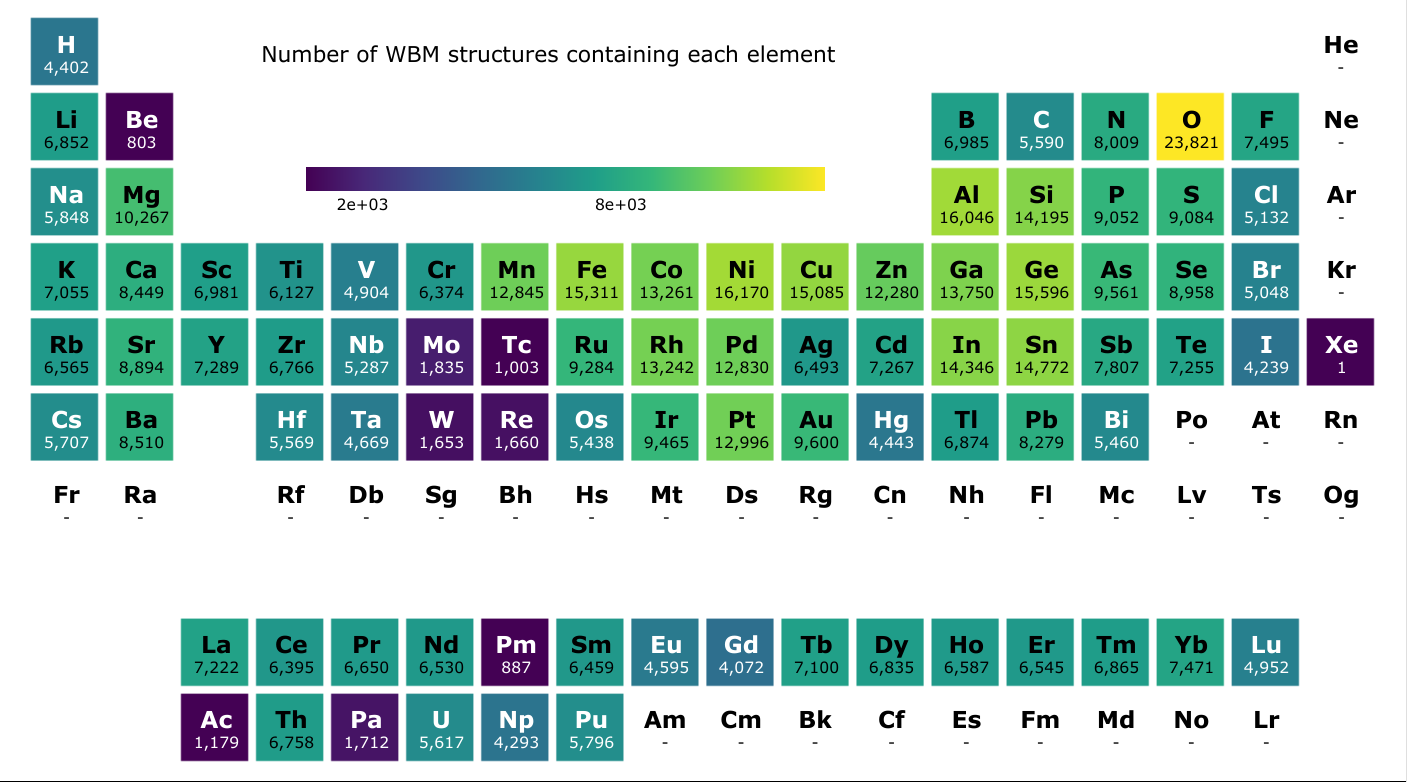}
    \caption{WBM test set element occurrence}
    \label{fig:wbm-element-counts-by-occurrence}
  \end{subfigure}
  \caption{
    The number of structures containing a given element in the MP training set and WBM test set \cite{wang_predicting_2021}.
    The WBM test set in relative terms contains noticeably fewer oxides than MP (and, by extension, MPtrj) with just 11\% rather than 53\% of structures containing oxygen.
    Made with pymatviz \cite{riebesell_pymatviz_2022}.
  }
  \label{fig:element-counts-by-occurrence}
\end{figure*}

\begin{figure*}
  \centering
  \begin{subfigure}[b]{0.9\linewidth}
    \includegraphics[width=\linewidth]{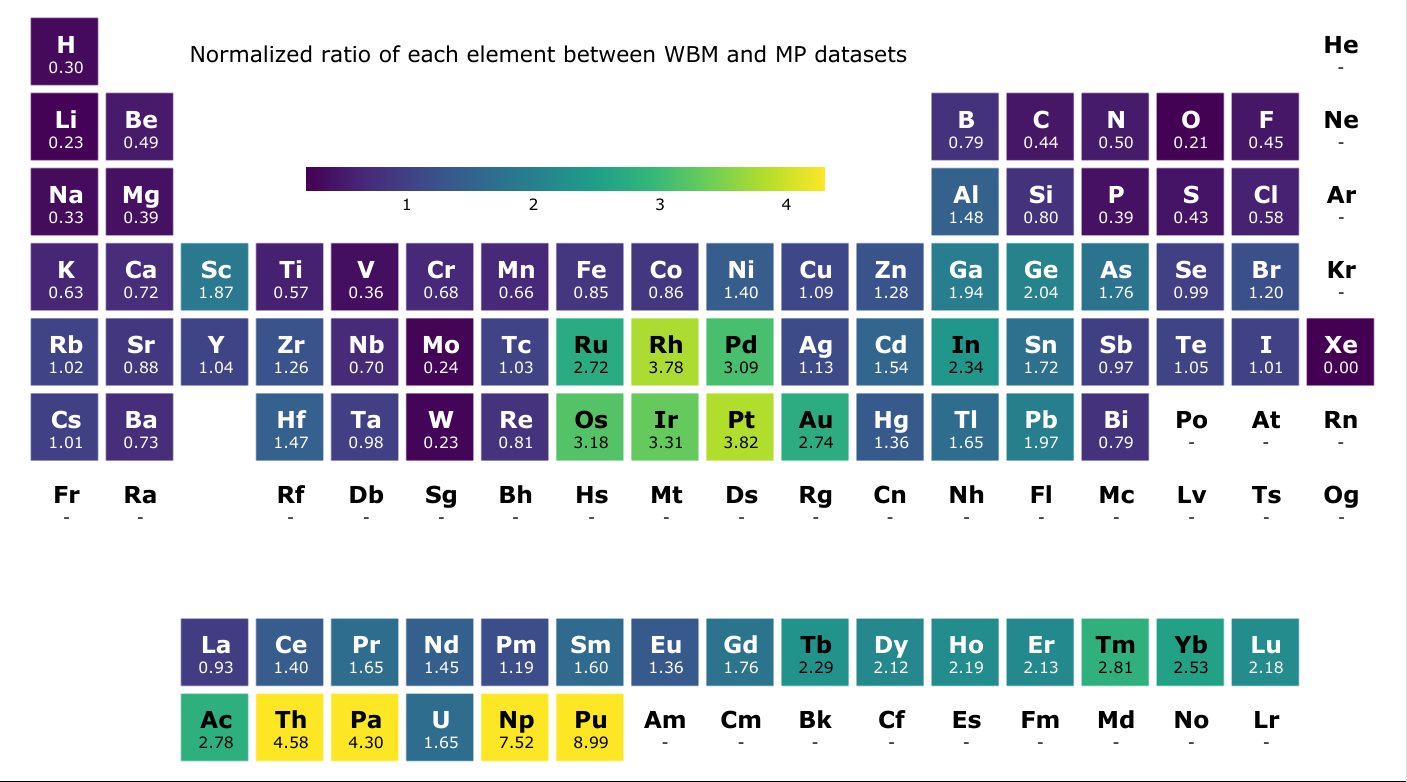}
    \caption{Normalized ratio of elements in WBM dataset to MP.}
    \label{fig:wbm-mp-ratio-element-counts-by-occurrence}
  \end{subfigure}
  \begin{subfigure}[b]{0.9\linewidth}
    \includegraphics[width=\linewidth]{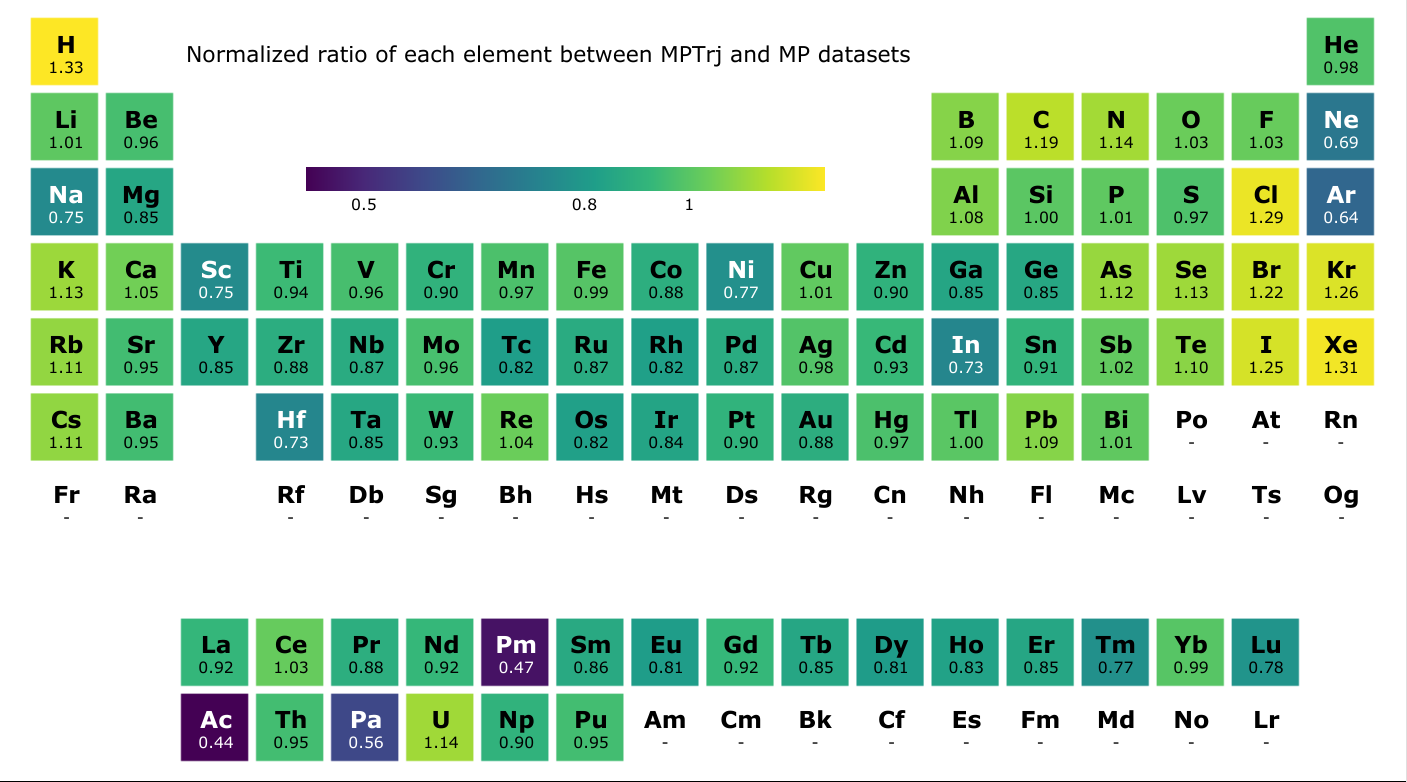}
    \caption{Normalized ratio of elements in MPtrj dataset to MP.}
    \label{fig:mp-trj-mp-ratio-element-counts-by-occurrence}
  \end{subfigure}
  \caption{
    \textbf{\subref*{fig:wbm-mp-ratio-element-counts-by-occurrence})} shows the ratio of elements in the WBM test set to the MP training set. The figure makes clear that WBM explores a distinct chemical space to MP with noble-transition metals, post-transition metals, lanthanides, actinides and metalloids seen more frequently in WBM than MP.
    Similarly, \textbf{\subref*{fig:mp-trj-mp-ratio-element-counts-by-occurrence})} shows the ratio of elements in the MPtrj dataset to the MP training set.
    We note a slight overabundance of structures containing hydrogen and halides, indicating that more frames were selected from structures containing these elements which might correlate with the number of ionic steps to find their ground states.
  }
  \label{fig:element-counts-ratio-by-occurrence}
\end{figure*}

\begin{figure*}
  \centering
\includegraphics[width=0.49\linewidth]{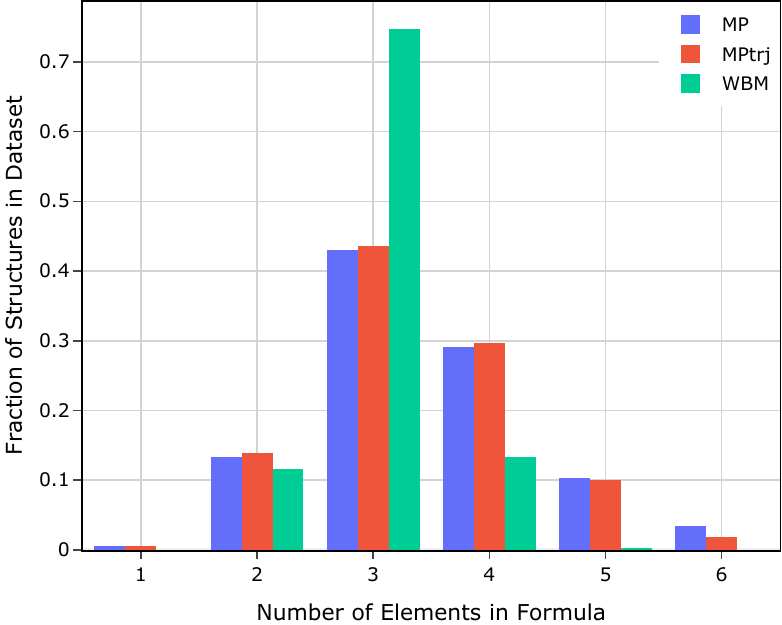}
  \caption{
    Distributions of unique elements per structure in MP, MPtrj and the WBM test set.
    The bar heights are normalized by the total number of structures in each dataset.
    WBM is dominated by ternary phases making up 75\% of the dataset followed by about 13\% for quaternaries and 12\% for binaries.
    MP has a more even distribution, in particular with more than double the relative share of quaternary phases and a significant number of quinternaries which are almost absent from WBM.
    Not shown in this plot for visual clarity are 3\% of MP structures containing more than 5 elements (up to 9).
    We also include MPtrj in this plot to show a slight drop in the relative abundance of quinternaries and higher phases vs MP ground states.
    This may be due to a poor choice of convergence criteria in early MP relaxation workflows that scaled with the size of the structure (see \texttt{EDIFF\_PER\_ATOM} parameter in \texttt{pymatgen} VASP input sets), resulting in unconverged large structures with short relaxation trajectories entering the database.
    Short relaxations would result in fewer frames of such structures selected for MPtrj.
    This assumes structures of higher arity correlate with larger structures.
  }
    \label{fig:mp-vs-mp-trj-vs-wbm-arity-hist}

\end{figure*}

\begin{figure*}
  \centering
  \begin{subfigure}[b]{0.45\linewidth}
    \includegraphics[width=\linewidth]{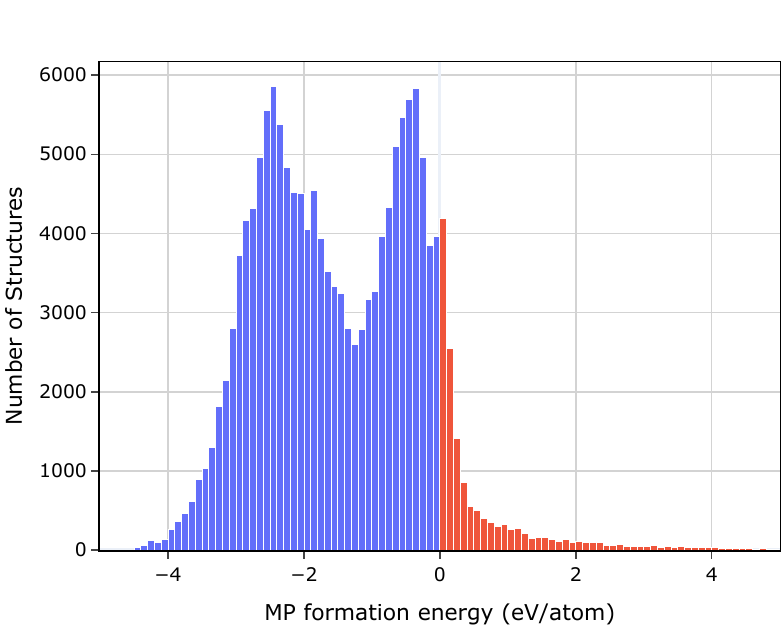}
    \caption{MP formation energy distribution}
    \label{fig:hist-mp-e-form-per-atom}
  \end{subfigure}
  \begin{subfigure}[b]{0.45\linewidth}
    \includegraphics[width=\linewidth]{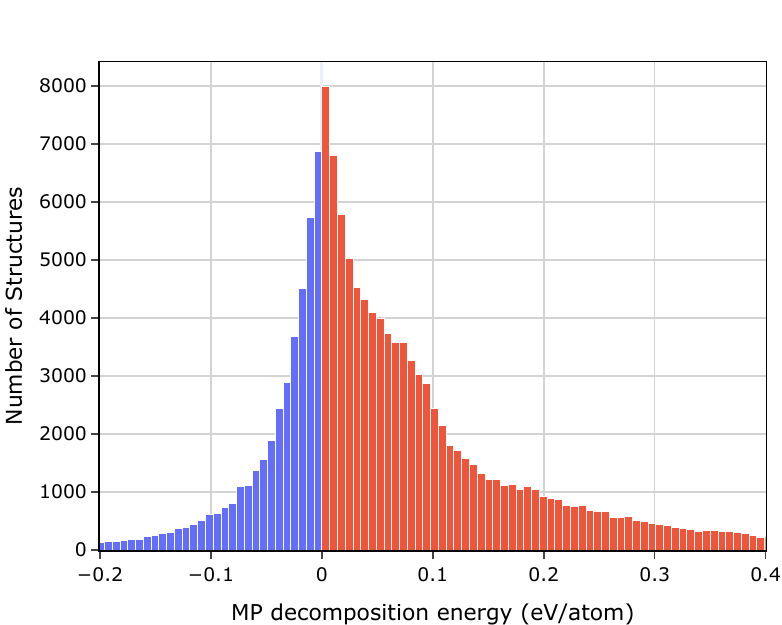}
    \caption{MP decomposition energy distribution}
    \label{fig:hist-mp-hull-dist}
  \end{subfigure}
  \begin{subfigure}[b]{0.45\linewidth}
    \includegraphics[width=\linewidth]{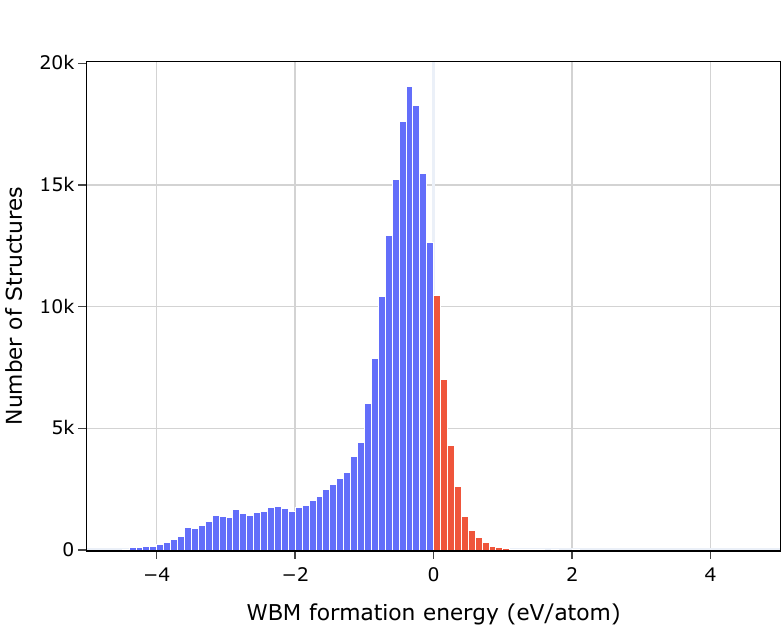}
    \caption{WBM formation energy distribution}
    \label{fig:hist-wbm-e-form-per-atom}
  \end{subfigure}
  \begin{subfigure}[b]{0.45\linewidth}
    \includegraphics[width=\linewidth]{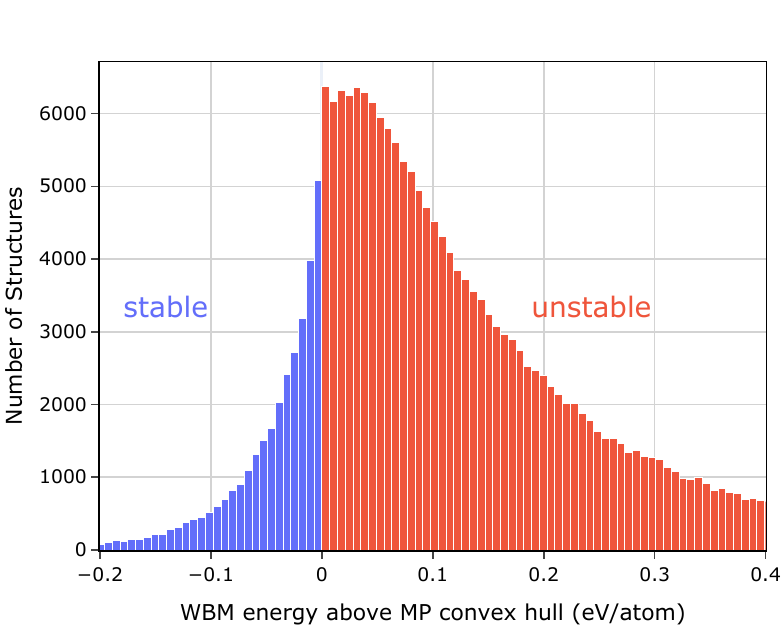}
    \caption{WBM convex hull distance distribution}
    \label{fig:hist-wbm-hull-dist}
  \end{subfigure}
  \caption{
    Distribution of formation energies, decomposition energy and convex hull distances for the MP training set and the WBM test set.
    In (\subref{fig:hist-mp-e-form-per-atom}) the bimodality in the MP formation energy distribution is due to the MP anion correction scheme \cite{wang_framework_2021} which significantly lowers some formation energies, especially for oxides. The decomposition energy shown in (\subref{fig:hist-mp-hull-dist}) calculated as defined in \cite{bartel_critical_2020}. We note that multiple materials with the same composition can have negative decomposition energies hence the number with negative decompositon energies is
    \simi43k compared to \simi35k that are on the hull. Looking at WBM in (\subref{fig:hist-wbm-e-form-per-atom}) and (\subref{fig:hist-wbm-hull-dist}), we see that the distribution of convex hull distances is much tigher due to the change of reference and therefore a more discriminative benchmarking task in terms of goodness-of-fit measures like the coefficient of determination.
  }
  \label{fig:wbm-energy-hists}
\end{figure*}

To give high-level insights into the MP training and WBM test set used in this work, we include element distributions for structures in both datasets (\cref{fig:element-counts-by-occurrence}).
To show how frame selection from MP structure relaxation affected relative elemental abundance between MP relaxed structures and the snapshots in MPtrj, \cref{fig:element-counts-ratio-by-occurrence} shows element occurrence ratios between MPtrj and MP.
Similarly, \cref{fig:mp-vs-mp-trj-vs-wbm-arity-hist} shows the elements-per-structure distribution of MP, MPtrj and WBM, normalized by dataset size.
The mode of all three datasets is 3, but WBM's share of ternary phases is noticeably more peaked than MP's, which includes small numbers of unary and senary phases.
\cref{fig:wbm-energy-hists} plots the distributions of formation energies, decomposition energies and convex hull distances (with respect to the convex hull spanned by MP materials only) for the MP and WBM data sets.
This plot not only gives insight into the nature of the dataset but also emphasizes the increased difficulty of stability vs formation energy prediction arising from the much narrower distribution of convex hull distances compared to the more spread-out formation energy distribution.
For WBM we see that the formation energy distribution exhibits much wider spread than the convex hull distance distribution, spanning almost \SI{10}{eV/atom} vs less than \SI{1}{eV/atom} spread in the hull distances.
This highlights why stability prediction is a much more challenging task than predicting energy of formation.
It requires correctly ranking the subtle energy differences between chemically similar compounds in the same chemical system rather than comparing a single material with the reference energies of its constituent elements.
DFT has been shown to significantly benefit from the systematic cancellation of errors between chemically similar systems when trying to identify the lowest-lying polymorph \cite{hautier_accuracy_2012}.
This beneficial cancellation has yet to be conclusively demonstrated for ML stability predictions.
So far, only the lack thereof has been shown in \cite{bartel_critical_2020} where they encountered a much more random error distribution among similar chemistries than simulations from first principles.

\begin{figure*}
  \centering
  \begin{subfigure}[b]{0.45\linewidth}
    \includegraphics[width=\linewidth]{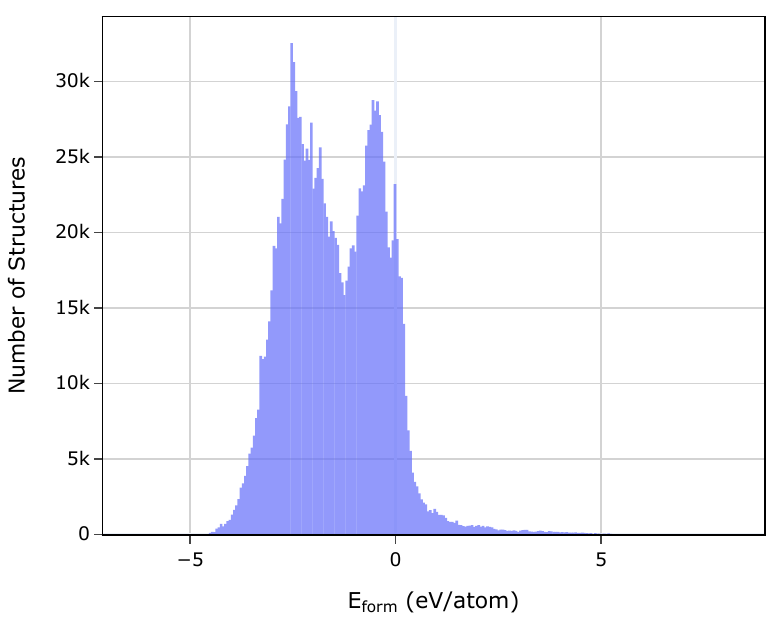}
    \caption{Distribution of formation energies.}
    \label{fig:mp-trj-e-form-hist}
  \end{subfigure}
  \begin{subfigure}[b]{0.45\linewidth}
    \includegraphics[width=\linewidth]{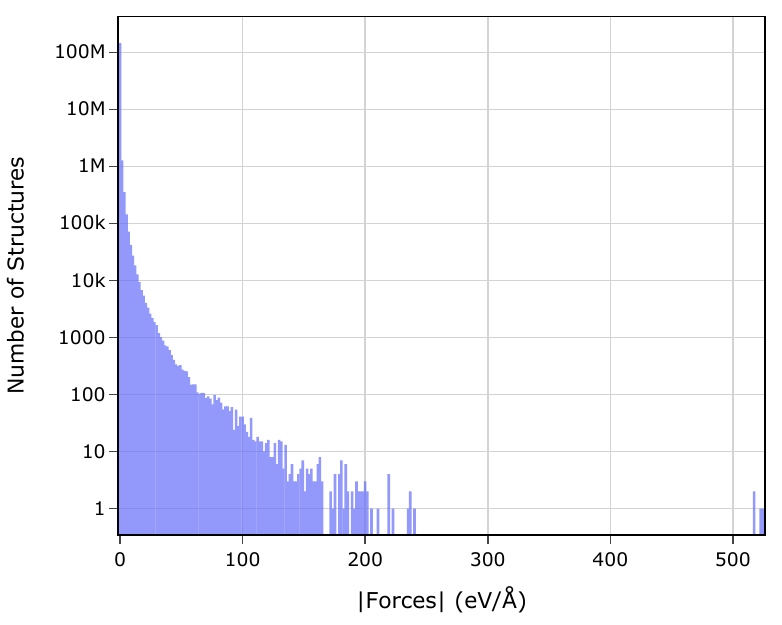}
    \caption{Distribution of forces.}
    \label{fig:mp-trj-forces-hist}
  \end{subfigure}
  \begin{subfigure}[b]{0.45\linewidth}
    \includegraphics[width=\linewidth]{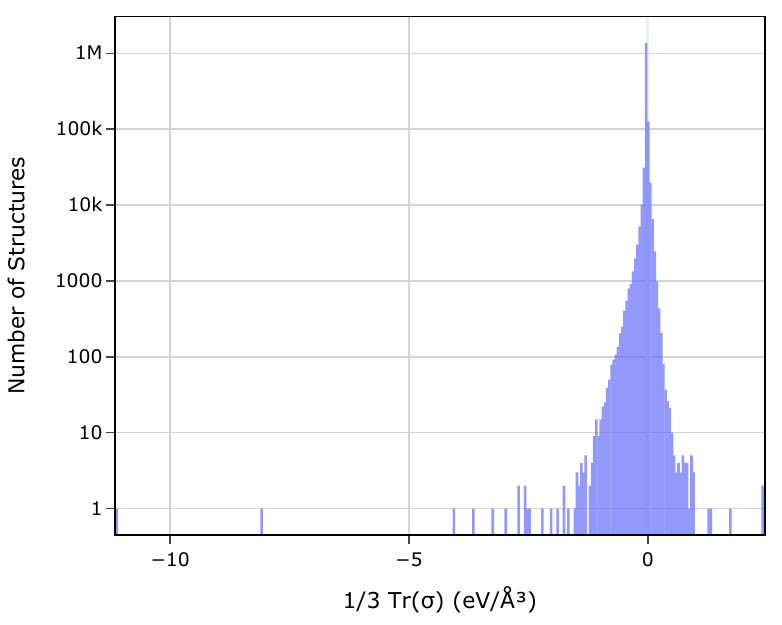}
    \caption{Distribution of stresses.}
    \label{fig:mp-trj-stresses-hist}
  \end{subfigure}
  \begin{subfigure}[b]{0.45\linewidth}
    \includegraphics[width=\linewidth]{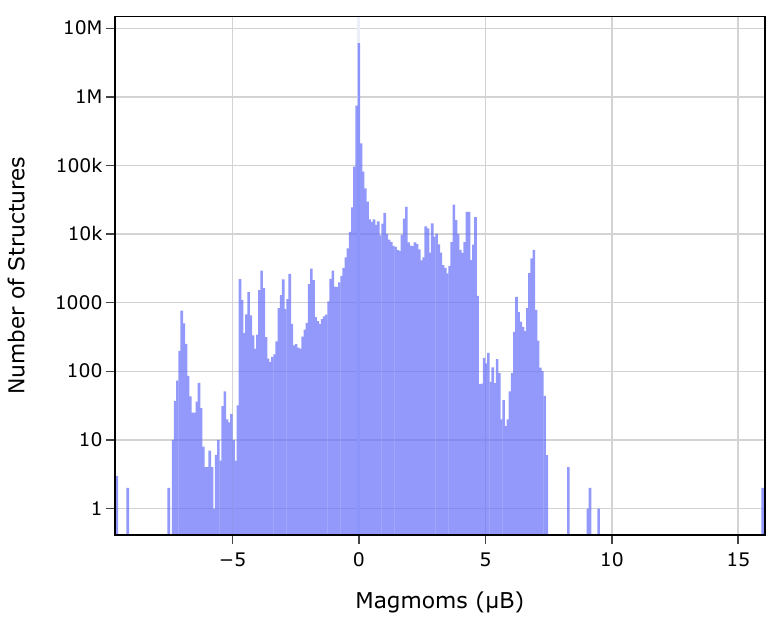}
    \caption{Distribution of magnetic moments.}
    \label{fig:mp-trj-magmoms-hist}
  \end{subfigure}
  \begin{subfigure}[b]{0.45\linewidth}
    \includegraphics[width=\linewidth]{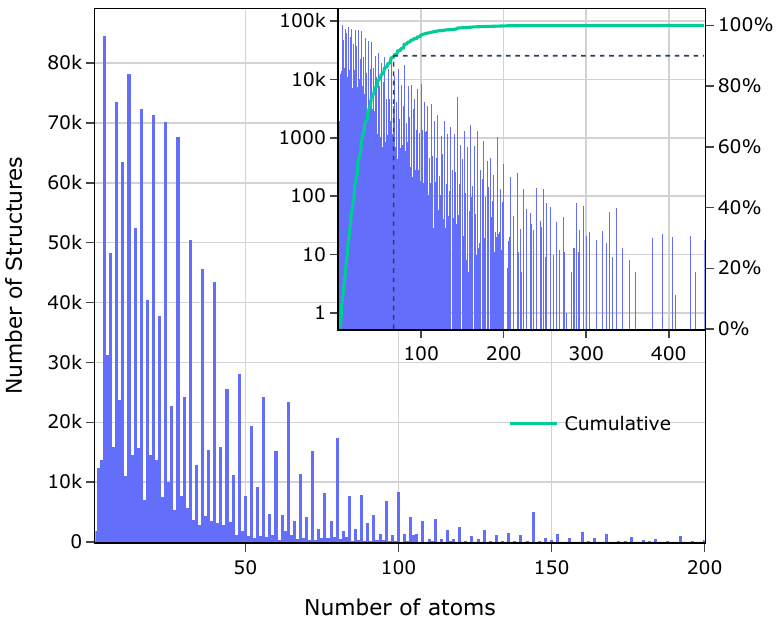}
    \caption{Distribution of the number of atomic sites.}
    \label{fig:mp-trj-n-sites-hist}
  \end{subfigure}
  \caption{
    Distribution of energies, forces, stresses, magnetic moments, and number of atoms in MPtrj.
    Comparing the distribution of formation energies to that of the MP dataset we see that the relative heights of the bimodal peaks is shifted as longer relaxation trajectories are seen on average for materials whose energies are adjusted by the anion correction scheme.
    The inset in (\subref{fig:mp-trj-n-sites-hist}) uses a log-scale to show the tail of the distribution. The green cumulative line in the inset shows that 82\% have less than 50 sites and 97\% of structures in MPtrj have less than 100 atoms.
  }
  \label{fig:mp-trj-hists}
\end{figure*}

\begin{figure*}[htbp!]
  \begin{subfigure}[b]{\linewidth}
    \centering
    \includegraphics[width=0.9\linewidth]{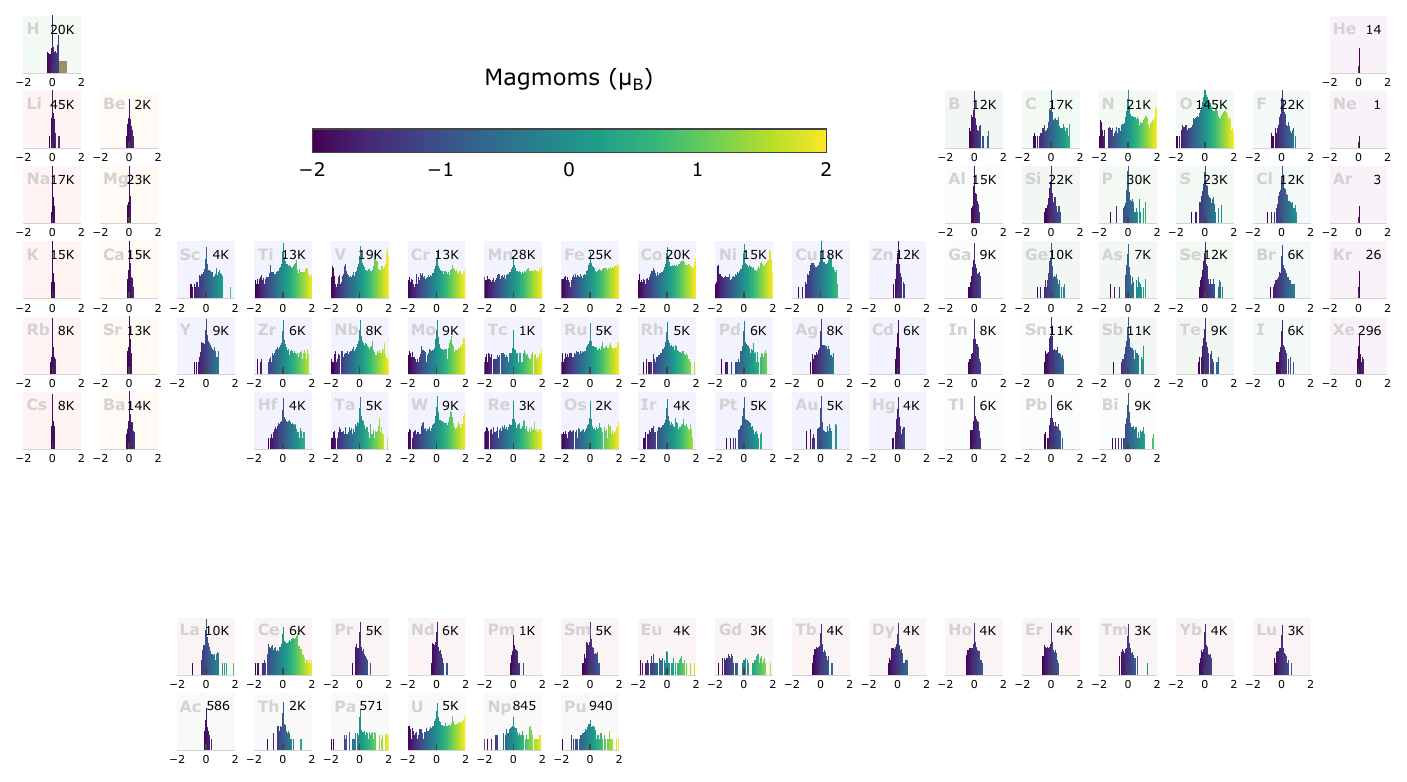}
    \caption{Element-wise MPtrj magnetic moment distributions}
    \label{fig:mp-trj-magmoms-ptable-hists}
  \end{subfigure}
  \begin{subfigure}[b]{\linewidth}
    \centering
    \includegraphics[width=0.9\linewidth]{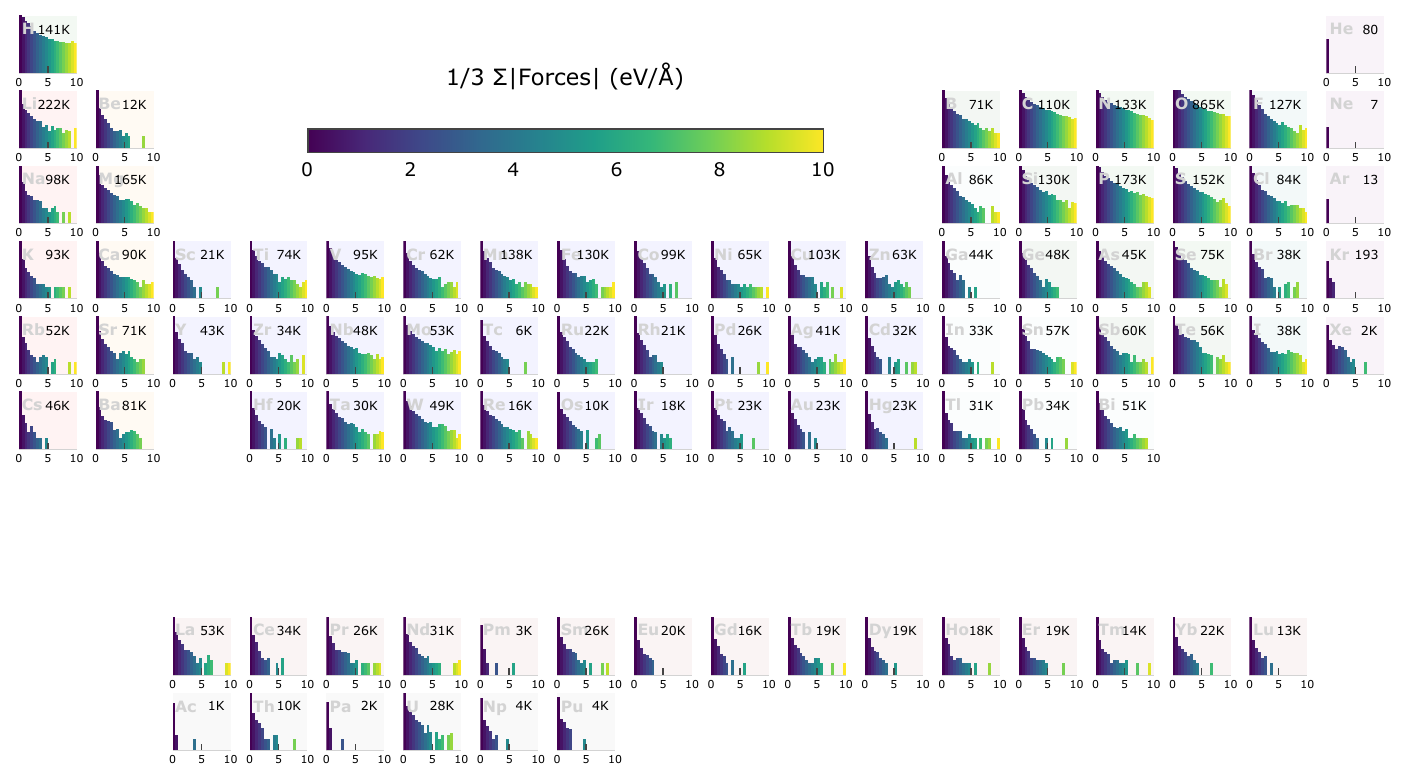}
    \caption{Element-wise MPtrj force distributions}
    \label{fig:mp-trj-forces-ptable-hists}
  \end{subfigure}
  \caption{
    Distribution of magnetic moments and forces for each element MPtrj. This data is used as training targets for all interatomic potentials in this work (only CHGNet uses the absolute value of magnetic moments as targets).
    The number in the top right corner of each element tile counts the number of target values for that element in all of MPtrj.
    $y$-axes are log-scaled to reveal the tail of high magnetic moments in some elements.
    \subref*{fig:mp-trj-magmoms-ptable-hists}) reveals rare erroneous data points in MPtrj.
    For instance, \ch{Cr} has a single-point calculation with a highly unphysical magnetic moment of \num{17}{$\mu_\text{B}$}.
    For visualization purposes, the $y$-axes are again log-scaled and distributions are truncated at \SI{10}{eV/\angstrom}.
    Oxygen has the largest outliers with mean absolute forces of up to \SI{160}{eV/\angstrom}.
  }
  \label{fig:mp-trj-ptable-hists}
\end{figure*}

Finally, Believing MPtrj to be an influential dataset for the near-term continued development of universal interatomic potentials, we plot histograms showing the distributions of target values for energies, forces, stresses and magnetic moments in \cref{fig:mp-trj-hists}.
The histogram in \cref{fig:mp-trj-n-sites-hist} shows the distribution of the number of sites in MPtrj structures. The inset displays the same histogram log-scaled y-axis as well as a cumulative line to show that 90\% of MPtrj structures contain fewer than 70 sites. \cref{fig:mp-trj-ptable-hists} shows the forces and magnetic moments for the MPTrj dataset projected onto different elements.

\end{document}